\RequirePackage{ifpdf}
\ifpdf 
\documentclass[pdftex]{sigma}
\else
\documentclass{sigma}
\fi

\numberwithin{equation}{section}

\begin{document}

\allowdisplaybreaks

\renewcommand{\thefootnote}{$\star$}

\renewcommand{\PaperNumber}{053}

\FirstPageHeading

\ShortArticleName{Higher Order Time Derivative Models}

\ArticleName{An Alternative Canonical Approach\\ to the Ghost Problem
in a Complexif\/ied Extension\\ of the Pais--Uhlenbeck Oscillator\footnote{This paper is a contribution to the Proceedings of the VIIth Workshop ``Quantum Physics with Non-Hermitian Operators''
   (June 29 -- July 11, 2008, Benasque, Spain). The full collection
is available at
\href{http://www.emis.de/journals/SIGMA/PHHQP2008.html}{http://www.emis.de/journals/SIGMA/PHHQP2008.html}}}

\Author{A. D\'ECTOR~$^{\dag}$, H.A. MORALES-T\'ECOTL~$^{\dag\ddag}$, L.F. URRUTIA~$^{\dag}$ and J.D. VERGARA~$^{\dag}$}

\AuthorNameForHeading{A.~D\'ector, H.A.~Morales-T\'ecotl, L.F.~Urrutia and J.D.~Vergara}

\Address{$^{\dag}$~Instituto de Ciencias Nucleares,
Universidad Nacional Aut\'onoma de M\'exico, \\
\hphantom{$^{\dag}$}~A. Postal 70-543, M\'exico D.F., M\'exico}
\EmailD{\href{mailto:dector@nucleares.unam.mx}{dector@nucleares.unam.mx}, \href{mailto:urrutia@nucleares.unam.mx}{urrutia@nucleares.unam.mx}, \href{mailto:vergara@nucleares.unam.mx}{vergara@nucleares.unam.mx}}

\Address{$^{\ddag}$~Departamento de F\'{\i}sica, Universidad Aut\'onoma Metropolitana Iztapalapa,\\
 \hphantom{$^{\ddag}$}~San Rafael Atlixco 186, Col. Vicentina, CP 09340, M\'exico D.F., M\'exico}
\EmailD{\href{mailto:hugo@xanum.uam.mx}{hugo@xanum.uam.mx}}

\ArticleDates{Received November 14, 2008, in f\/inal form April 22,
2009; Published online May 05, 2009}

\Abstract{Our purpose in this paper is to analyze the Pais--Uhlenbeck
(PU) oscillator using complex canonical transformations. We show
that starting from a Lagrangian approach we obtain a transformation
that makes the extended PU oscillator, with unequal frequencies, to
be equivalent to two standard second order oscillators which have
the original number of degrees of freedom. Such extension is
provided by adding a total time derivative to the PU Lagrangian
together with a complexif\/ication of the original variables further
subjected to reality conditions in order to maintain the required
number of degrees of freedom. The analysis is accomplished at both
the classical and quantum levels. Remarkably, at the quantum level
the negative norm states are eliminated, as well as the problems of
unbounded below energy and non-unitary time evolution. We illustrate
the idea of our approach by eliminating the negative norm states in
a complex oscillator. Next, we extend the procedure to the
Pais--Uhlenbeck oscillator. The corresponding quantum propagators are
calculated using Schwinger's quantum action principle. We also
discuss the equal frequency case at the classical level.}

\Keywords{quantum canonical transformations; higher order derivative models}

\Classification{70H15; 70H50; 81S10}

\renewcommand{\thefootnote}{\arabic{footnote}}
\setcounter{footnote}{0}

\section{Introduction}

Systems with higher order time derivatives (HOTD) have been studied
with increasing interest because they appear in many important
physical problems. In f\/irst place these systems were considered to
improve the divergent ultraviolet behavior of some quantum f\/ield
theories~\cite{Thirring:1950}. However, their energy turned out not
bounded from below \cite{Pais:1950za}, then involving ghosts and
making the theory non-unitary \cite{Heisenberg1957}. In spite of
such drawbacks, higher order derivative theories were studied to
learn on their improved renormalization properties. Even a
renormalizable higher order quantum gravity theory was advanced
along these lines \cite{Stelle:1976gc} and the unitarity of a
lattice form was studied in \cite{Tomboulis:1983sw}. See also
\cite{Hawking:2001yt,Moeller:2002vx,Rivelles:2003jd,
Antoniadis:2006pc,Codello:2006in,Berkovits:2006vc} and references
therein for other examples. As for dealing with ghosts, attempts can
be classif\/ied according to whether the approach is perturbative
\cite{Jaen:1986iz,Eliezer:1989cr,Cheng:2001du} or not
\cite{Simon:1990ic,Hawking:2001yt,Rivelles:2003jd}, but a def\/inite
answer is yet to be found.

Since single particle quantum mechanics can be seen, in the free
f\/ield or weak coupling limit, as a mini-superspace sector of quantum
f\/ield theory where the spatial degrees of freedom have been frozen,
it is suggestive to test new ideas on HOTD theories, by using
quantum mechanics as a~laboratory. The model example to do so is the
Pais--Uhlenbeck (PU) oscillator~\cite{Pais:1950za}, which consists of
a~one dimensional harmonic oscillator Lagrangian plus a term
quadratic in acceleration. A~standard quantum treatment of its
degrees of freedom gives rise to a spectrum not bounded from below
since it consists of the dif\/ference of two quantum harmonic
oscillators spectra \cite{Pais:1950za}. An interesting solution to
such dif\/f\/iculty has been recently proposed in
\cite{Bender:2007wu,Bender:2008vh1,Bender:2008vh} (see also
\cite{Smilga:2004cy,Smilga:2005gb,Smilga1,Smilga:2008hn} for a
dif\/ferent approach) by f\/inding a quantum transformation which gives
its Hamiltonian a non-Hermitian, $\cal PT$ symmetric form. In this
approach a technique has been developed to obtain a physical inner
product and thus a Hilbert space \cite{Bender:2007nj}.

The aim of this paper is to show that it is possible to tackle the
problem of quantizing an extension of the PU oscillator within a
Lagrangian and a canonical ormulation, using complex canonical
transformations and non-Hermitian variables together with reality
conditions. This approach is motivated by a previous proposal to
build complex canonical variables for non-perturbative quantum
canonical general relativity \cite{Ashtekar:1991hf,Ashtekar:1992ne},
which achieved the nontrivial task of making polynomial the
constraints of general relativity, whereas the so called reality
conditions to be fulf\/illed by the non-Hermitian variables were
proposed to determine the inner product of Hilbert space. For
canonical general relativity such reality conditions imply the
metric of space and its rate of change are real quantities. Some
ef\/fort was devoted to exploit these complex variables (see for
instance \cite{Thiemann:1995ug,Ashtekar:1995qw,Montesinos:1999qc})
however the use of real variables allowed f\/inally for important
progress \cite{Rovelli:2004tv,Thiemann:2007zz}.

The paper is organized as follows. In Section~\ref{section2} we review the essential
aspects of the general construction with special emphasis on the
boundary conditions. In Section~\ref{section3} we use as a toy model a complexif\/ied
harmonic oscillator \cite{Ashtekar:1992ne}. Although this is not a
higher-order derivative problem it exhibits negative norm states. We
show that using the reality conditions we can eliminate them.
Furthermore, we evaluate the Green's function of the system using
the Schwinger quantum action principle and the path integral method,
f\/inding that both of these procedures lead to the same result. In
Section~\ref{section4} we combine the results of the previous sections in order to
 make a f\/irst attempt to extend the formalism to HOTD theories,
 successfully taking as an example the Pais--Uhlenbeck oscillator,
which can be treated in a similar way as the complexif\/ied harmonic
oscillator. In addition we perform the canonical analysis in an
extended phase space in order to compare the cases of PU with equal
and unequal frequencies. We observe that in the case of equal
frequencies, the Lagrangian transformation induces a non canonical
one which, at the quantum level, will lead to a non diagonalizable
Hamiltonian. Finally, concluding remarks are presented in Section~\ref{section5}.
Unless otherwise stated we use units in which $\hbar=1$.

\section{General considerations}\label{section2}

Let us consider a system whose Lagrangian contains up to the $n$-th order time derivative in a~quadratic form. Written in normal coordinates this Lagrangian will have the form,
\begin{equation}\label{HOL}
 L(x,\dot{x},\dots,x^{(n)})=\sum_i \alpha_i \big(x^{(i)}\big)^2- V(x), \qquad i=1,\dots, n,
\end{equation}
with $x^{(k)}= \frac{d^k x}{dt^k}$ and the $\alpha_i\in \mathbb{R}$ being constants. We are assuming that under the
transformation to normal coordinates we diagonalize the kinetic term and that the potential only depends on the
coordinates. This Lagrangian yields an equation of motion of order $(2n)$
\begin{gather*}
\sum_{k=0}^{n} (-1)^k\frac{d^k\mbox{}}{dt^k} \frac{\partial L}{\partial x^{(k)}} = 0 .
\end{gather*}
Assuming $x$ and hence $L$ to be real leads to the well known
problems of the higher-order derivative models. So, as a f\/irst step
in our extension we consider the analytic continuation of the
variable $x$ to the complex plane. Next we assume that the
Lagrangian (\ref{HOL}) plus a total time derivative will be real.
This certainly does not alter the classical equations of motion, but
modif\/ies the Hamiltonian formulation by the addition of new terms in
the momenta. Now, the essential points of our procedure are the following two i)
we assume that there exists a nonlocal transformation from the Lagrangian
(\ref{HOL}) plus a total time derivative to a second order real
Lagrangian with $n$ independent conf\/iguration variables $\xi^i$.
Specif\/ically we assume the complexif\/ied description can be related
to an alternative one in terms of a real Lagrangian
$L_{\xi}\big(\xi^1,\dot{\xi}^1,\dots,\xi^n,\dot{\xi}^n\big)$ through
\begin{gather}
L\big(x,\dot{x},\dots,x^{(n)}\big) + \frac{df}{dt} = L_{\xi}\big(\xi^1,\dot{\xi}^1,\dots,\xi^n,\dot{\xi}^n\big) ,\label{lamod}\\
f = f\big(x,\dots,x^{(n)}\big) ,\label{gefunc}\\
\xi^i = \xi^i\big(x,\dots,x^{(n)}\big), \qquad i=1,\dots, n . \label{xix}
\end{gather}
Here $\xi^i$, $i=1,\dots,n$ are real coordinates and only their f\/irst
order time derivatives $\dot{\xi}^i$ enter in $L_{\xi}$ in a
quadratic form. In general $f$ will be complex. ii) The second point
is that the transformation (\ref{xix}) will impose reality
conditions on the $x^{(k)}$, with $k=0,\dots,n$. This implies, in
particular, that terms in the potential of higher order than
quadratic must be real functions of the variables, i.e., under the
analytic continuation, we must consider only the real part of the
interaction, for example powers like~$x^4$ should be replaced by
$(xx^\dag)^2$. This is not necessary for the quadratic terms, since
our transformation incorporates them properly. We must notice that
strictly speaking, our system is dif\/ferent from the original one,
since we have analytically continued the $x$ variable. However, the
reality conditions allow us to construct a properly well def\/ined
problem, with the same number of the original degrees of freedom
together with the same classical equation for the original variable~$x$.

\subsection{Canonical transformation and boundary conditions}\label{section2.1}

One of the interesting properties of our nonlocal transformation
(\ref{xix}) is that in the Hamiltonian--Ostrogradsky formalism
\cite{OSTROGRADSKI}, this transformation corresponds to a local
canonical transformation. To see this, we observe that in this
formalism the corresponding phase space will be of dimension~$2n$,
so the conf\/iguration space is extended to $Q_0=x$, $Q_1= \dot
x,\dots, Q_{n-1}=x^{(n-1)}$, together with the momenta
\begin{gather}\label{momen}
\Pi_i=\sum_{j=0}^{n-i-1} \left(-\frac{d}{dt}\right)^j\frac{\partial
L}{\partial q^{(i+j+1)}}, \qquad i=0,\dots, n-1.
\end{gather}
Thus, the transformation (\ref{xix}) only depends on the indicated phase space variables,
\begin{equation}\label{mxix}
\xi^i\big(x,\dots,x^{(n)}\big)=\xi^i(Q_0, Q_1,\dots,Q_{n-1}, \Pi_{n-1}),
\end{equation}
in which only the last momentum $\Pi_{n-1}$ appears. We also have
the corresponding transformation for the momenta,
\begin{equation}\label{momentrans}
P_i=P_i(Q_0, Q_1,\dots,Q_{n-1},\Pi_{0},\dots , \Pi_{n-1}),
\end{equation}
given directly from the generating function (\ref{gefunc}). The
transformations (\ref{mxix}), (\ref{momentrans}) will be canonical in
an extended phase space because the Lagrangians $L$ and $L_\xi$
dif\/fer only by a total derivative that depends on the coordinates and momenta. We
must remark that the transformations (\ref{mxix}), (\ref{momentrans})
should be complex, because an analytic continuation of the original
variables was required. However, these transformations def\/ine exactly the reality
conditions, since we enforce that the $2n$ variables $(\xi^i,P_i)$
will be real and, in consequence, also def\/ine the integration contour
that must be used to quantize the system.

To build the explicit form of the canonical transformation that we are using, we consider the equivalent of
the expression (\ref{lamod}) in Hamiltonian form,
\begin{equation}\label{canoham}
\Pi_i \dot Q^i - \mathcal{H}(Q,\Pi)+ \frac{df(Q,\Pi)}{dt}=P_i \dot \xi^i -H(\xi , P).
\end{equation}
From the comparison of independent terms we find
\begin{equation}\label{genera}
\frac{\partial f}{\partial Q^i}=P_j \frac{\partial \xi^j}{\partial Q^i} -\Pi_i, \qquad
\frac{\partial f}{\partial \Pi_i}= P_j \frac{\partial \xi^j}{\partial \Pi_i}, \qquad
\mathcal{H}=H.
\end{equation}
These equations prescribe the $2n$ partial derivatives of the generating function $f(Q,\Pi)$ and the Hamiltonians are equal
due to the fact that the transformation is time independent.

Now, to analyze the boundary conditions that are imposed by our
complex canonical transformation, we make the variation of the left
hand side of expression (\ref{canoham}) and we observe that the
boundary condition is associated with the total derivative
\begin{equation*}
\frac{d}{dt}\left(\left(\Pi_i + \frac{\partial f}{\partial
Q^i}\right) \delta Q^i +\frac{\partial f }{\partial \Pi_i} \delta
\Pi_i\right),
\end{equation*}
which dictates the appropriate combinations of the variations $\delta
Q^i$ and $\delta \Pi_i$ to be f\/ixed at the boundaries. Using
 the equations~(\ref{genera}) we can show that the above expression is indeed
 $\frac{d}{dt}\big(P_i\delta \xi^i \big)$, in terms of
the variation of the~$\xi^i$. So, the variables that we need to f\/ix
on the boundary are those functions of~$Q^i$ and~$P_i$ def\/ined by
the~$\xi^i$'s.

\section{Reality conditions and the complex oscillator}\label{section3}

Let us start by reviewing the example of a modif\/ied harmonic oscillator subject to a complex canonical
transformation \cite{Ashtekar:1991hf,Ashtekar:1992ne}. This has also been studied in a form that exploits
the ${\cal PT}$ symmetry in the framework of the non-Hermitian Hamiltonian approach
\cite{Bender:2007nj,Swanson,Jones,Ivanov:2007me}. The strategy is based on the observation that the complex Lagrangian
\begin{gather*}
L_{\rm C} = \frac{\dot{q}^2}{2}-\frac{q^2}{2}- i\epsilon q\dot{q} ,
\end{gather*}
where $\epsilon$ is a real parameter, becomes the one corresponding to the harmonic oscillator after adding
to it the total time derivative $\frac{df_{\rm C}}{dt}$, $f_{\rm C} = i\epsilon \frac{q^2}{2}$. In the Hamiltonian description
the canonical momentum is
\begin{gather*}
p = \frac{\partial L_{\rm C}}{\partial \dot{q}}=\dot{q}-i\epsilon q,
\end{gather*}
so $p\in {\mathbb C}$ and should be quantized as a non-Hermitian
operator.

\subsection{Hilbert space and reality conditions}\label{section3.1}

The quantum Hamiltonian becomes
\begin{equation}
\label{Hmod}
\hat{H}_{\rm C}=\frac{\hat{p}^{2}}{2}+\frac{\hat{q}^2}{2}-\frac{1}{2} \epsilon^{2}\hat{q}^{2}+
\frac{i\epsilon}{2}\left\{\hat{p} ,\hat{q}\right\} ,
\end{equation}
where $\{\hat{p},\hat{q}\}=\hat{p}\hat{q}+\hat{q}\hat{p}$ so that a
symmetric ordering is selected. Upon the canonical transformation
$\hat p=\hat P-i\epsilon \hat Q$, $\hat q=\hat Q$, with $\hat
P=\frac{d\hat Q}{dt}$, the Hamiltonian (\ref{Hmod}) becomes the
standard one of the harmonic oscillator $\hat H_{\rm HO}= \frac{\hat
P^2}{2}+\frac{\hat Q^2}{2}$, which is Hermitian whenever $\hat Q$
and $\hat P$ are.

The coordinate representation
\begin{gather*}
\hat{q}\psi(q)=q\psi(q) , \qquad \hat{p}\psi(q)=-i\frac{d\psi}{dq}(q) ,
\end{gather*}
leads to a non-Hermitian form of the Hamiltonian (\ref{Hmod}) with
the usual scalar product\footnote{Using $\hat{p}=-i\frac{d\mbox{}}{dq} + i\epsilon q$ would yield the usual Hamiltonian for the harmonic oscillator which, however, requires a~modif\/ied inner product to be Hermitian.}. The corresponding Schr\"odinger
equation becomes
\begin{equation*}
\psi''( q)-2\epsilon q \psi'( q)-\left((1-\epsilon^{2})q^2+\epsilon-2 E\right)\psi( q)=0.
\end{equation*}
Hence the eigenvalue problem for (\ref{Hmod}) can be related to that
of the harmonic oscillator by using the wave functions of the latter
$\varphi_n=N_n\mathrm{e}^{-\frac{q^2}{2}}H_n(q)$ and def\/ining
\begin{equation} \label{psin}
\psi_{n}( q):={\rm e}^{\frac{\epsilon q^{2}}{2}}\varphi_n ,\qquad E_{n}=n+\tfrac{1}{2} ,\qquad n=0,1,2\dots .
\end{equation}
A tedious but otherwise direct calculation shows that the eigenfunctions $\psi_n$ do not have positive norm in the
Hilbert space ${\cal H}_0=L^2({\mathbb R},dq)$. However, the classical role of $f_C$ as the generator of a canonical
transformation motivates the introduction of a quantum transformation such that (see for example~\cite{Anderson:1993ia,Anderson:1993im} for a general discussion)
\begin{equation}\label{CTho}
\hat{{H}}_{{\rm HO}}={\rm e}^{-\epsilon \frac{q^2}{2}}\hat{{H}}_{\rm C} {\rm e}^{\epsilon \frac{q^2}{2}} .
\end{equation}
This non-unitary canonical transformation changes the measure $dq$ to $d\mu=\mathrm{e}^{-\epsilon q^2} dq $. Now the
Hilbert space ${\cal H}=L^2({\mathbb R},d\mu)$ ensures (\ref{psin}) have positive norm since
\begin{gather*}
\langle n|m\rangle_{\mu} := \int dq \mathrm{e}^{-\epsilon q^2}\psi_n^{\ast}\psi_m
 = \int dq \varphi_n^{\ast}\varphi_m
=: \langle n|m\rangle_q =\delta_{mn} ,
\end{gather*}
where subscripts denote the appropriate measure for the correspondent Hilbert space.
Moreover, the reality conditions
\begin{equation}\label{realitycon}
\hat{q}^{\dag} = \hat{q} ,\qquad \hat{p}^{\dag} = \hat{p} + 2i\epsilon \hat{q}
\end{equation}
are automatically implemented in $\cal H$, where (\ref{Hmod}) is Hermitian~\cite{Ashtekar:1991hf}.

It is illuminating to consider the Green's functions corresponding, respectively, to~$\hat{H}_{\rm C}$ \linebreak and~$\hat{H}_{\rm HO}$. We recall they can be def\/ined as
\begin{gather}
G^{\epsilon}_{\rm C}(q_2,q_1,E) := \langle q_2 |\frac{1}{E-\hat{H}_{\rm C}} |q_1\rangle , \label{GC}\\
G_{\rm HO}(q_2,q_1,E) := \langle q_2 |\frac{1}{E-\hat{H}_{\rm HO}} |q_1\rangle . \label{GHO}
\end{gather}
Now on account of (\ref{CTho}), we can rearrange (\ref{GC}) to relate it with (\ref{GHO}) as follows
\begin{gather}
G^{\epsilon}_{\rm C}(q_2,q_1,E) = \langle q_2|\frac{1}{E-\hat{\cal O}^{-1}_{\epsilon}\hat{H}_{\rm HO}\hat{\cal O}_{\epsilon}}|q_1\rangle
 = \langle q_2|{\hat{\cal O}^{-1}_{\epsilon}} \frac{1}{E-\hat{H}_{\rm HO}}{\hat{\cal O}_{\epsilon}}|q_1\rangle\nonumber\\
\phantom{G^{\epsilon}_{\rm C}(q_2,q_1,E)}{} = \mathrm{e}^{+\epsilon(\frac{q^2_2}{2}-\frac{q^2_1}{2})} G_{\rm HO}(q_2,q_1,E), \qquad \hat{\cal O}_{\epsilon}:= \mathrm{e}^{-\epsilon\frac{\hat{q}^2}{2}} , \label{25}
\end{gather}
where in the f\/inal line we use that $\hat{q}$ acts diagonally on
the basis $|q\rangle$ and its dual. Thus the standard harmonic
oscillator Green's function, $G_{\rm HO}(q_2,q_1,E)$, is obtained from
that corresponding to the complex oscillator,
$G^{\epsilon}_{\rm C}(q_2,q_1,E)$, upon multiplying the former by
$\mathrm{e}^{\epsilon\big(\frac{q^2_2}{2}-\frac{q^2_1}{2}\big)}$.

\subsection{The Schwinger quantum action principle}\label{section3.2}

We can use Schwinger's quantum action principle to derive some dynamical properties of the system \cite{Schwinger2001}. For instance, we can calculate the propagator $\langle q_{2},t_{2}|q_{1},t_{1}\rangle$. We will work in the Heisenberg picture, where we adopt the notation
\begin{alignat}{3}
& \langle q_{2},t_{2}|q(t_{2})=q_{2}\langle q_{2},t_{2}|,\qquad && q(t_{1})|q_{1},t_{1}\rangle=q_{1}|q_{1},t_{1}\rangle, & \label{3.1}\\
& \langle p^{\ast}_{2},t_{2}|p^{\dagger}(t_{2})=p^{\ast}_{2}\langle p^{\ast}_{2},t_{2}|,\qquad && p(t_{1})|p_{1},t_{1}\rangle=p_{1}|p_{1},t_{1}\rangle, &\label{3.2}
\end{alignat}

To achieve our purpose, we f\/irst calculate the total variation of $\langle q_{2},t_{2}|q_{1},t_{1}\rangle$, given by
\begin{gather*}
\delta \langle q_{2},t_{2}|q_{1},t_{1}\rangle=i\langle q_{2},t_{2}|G_{q_{2}}-G_{q_{1}}-H\delta T|q_{1},t_{1}\rangle ,
\end{gather*}
where $T=t_{2}-t_{1}$, and the generators $G_{q_{i}}$ are given by
\begin{gather*}
G_{q_{2}}=p(t_{2})\delta q_{2} ,\qquad G_{q_{1}}=p(t_{1})\delta q_{1} .
\end{gather*}
Starting from the Hamiltonian operator (\ref{Hmod}), we get the following Heisenberg operator equations
\begin{gather*}
\dot{q}(t) = p(t)+i\epsilon q(t) ,\nonumber\\
\dot{p}(t) = -(1-\epsilon^{2})q(t)-i \epsilon p(t) .
\end{gather*}
Fixing the f\/inal and initial conditions at $t_{2}$ and $t_{1}$ as $q\left(t_{2}\right)$ and $q\left(t_{1}\right)$, we obtain the solutions
\begin{gather}
q(t) = A \cos(t)+B \sin(t) ,\label{solucionq}\\
p(t) = (B-i\epsilon A) \cos(t)-(i \epsilon B+A) \sin(t) ,\label{solucionp}
\end{gather}
where the coef\/f\/icients $A$ and $B$ are given by
\begin{gather*}
A = \frac{1}{\sin\left(T\right)}\left(q(t_{1})\sin\left(t_{2}\right)-q(t_{2})\sin\left(t_{1}\right)\right) ,\\
B = \frac{1}{\sin\left(T\right)}\left(q(t_{2})\cos\left(t_{1}\right)-q(t_{1})\cos\left(t_{2}\right)\right) .
\end{gather*}
In this way, we can express $p(t_{2})$ and $p(t_{1})$ in terms of the initial and f\/inal conditions $q(t_{1})$ and~$q(t_{2})$, getting
\begin{gather*}
p(t_{1}) = \frac{1}{\sin(T)}\bigl(-(i \epsilon \sin(T)+\cos(T))q(t_{1})+q(t_{2})\bigr) ,\\ 
p(t_{2}) = \frac{1}{\sin(T)}(-(i\epsilon \sin(T)-\cos(T))q(t_{2})-q(t_{1})) .
\end{gather*}
With this relations, one can also calculate the commutator
\begin{gather*}
\left[q(t_{1}),q(t_{2})\right]=i\sin\left(T\right).
\end{gather*}
Using all of the above, we can write the Hamiltonian as
\begin{gather*}
H = \frac{1}{2 \sin^{2}(T)}\left(q^{2}(t_{1})+q^{2}(t_{2})-2 \cos(T)q(t_{2}) q(t_{1})\right)-\frac{i}{2}\frac{\cos(T)}{\sin(T)} .
\end{gather*}
Finally, from the equations (\ref{3.1}), (\ref{3.2}) we get:
\begin{gather*}
\delta\langle q_{2},t_{2}|q_{1},t_{1}\rangle=i\langle q_{2},t_{2}|\Biggl\{\biggl\{\frac{1}{\sin(T)}\left(\cos(T)q_{2}-q_{1}\right)-i\epsilon q_{2}\biggr\}\delta q_{2}\nonumber\\
\phantom{\delta\langle q_{2},t_{2}|q_{1},t_{1}\rangle=}{} +\left\{\frac{1}{\sin(T)}(\cos(T)q_{1}-q_{2})+i\epsilon q_{1}\right\} \delta q_{1}\nonumber\\
\phantom{\delta\langle q_{2},t_{2}|q_{1},t_{1}\rangle=}{} -\biggl\{\frac{1}{2 \sin^{2}(T)}\left(q_{1}^{2}+q_{2}^{2}-2 \cos(T)q_{2} q_{1}\right)-\frac{i}{2}\frac{\cos(T)}{\sin(T)}\biggr\}\delta T \Biggr\}|q_{1},t_{1}\rangle \nonumber\\
\phantom{\delta\langle q_{2},t_{2}|q_{1},t_{1}\rangle}{} =i\langle q_{2},t_{2}|\delta\Biggl\{\frac{1}{2\sin(T)}\left((q_{2}^{2}+q_{1}^{2})\cos(T)-2q_{2}q_{1}\right)\nonumber\\
\phantom{\delta\langle q_{2},t_{2}|q_{1},t_{1}\rangle=}{} -\frac{i\epsilon}{2}(q_{2}^{2}-q_{1}^{2})-i\ln\left(\frac{1}{\sqrt{\sin(T)}}\right)\Biggr\}|q_{1},t_{1}\rangle .
\end{gather*}
Integrating we obtain
\begin{gather}
\langle q_{2},t_{2}|q_{1},t_{1}\rangle =\frac{1}{\sqrt{2\pi i\sin(T)}}
\exp\left\{\frac{i}{2 \sin(T)}\bigl\{(q_{2}^{2}+q_{1}^{2})\cos(T)-2q_{2}q_{1}\bigr\}
\right\}\nonumber\\
\phantom{\langle q_{2},t_{2}|q_{1},t_{1}\rangle =}{}\times
\exp\left\{\frac{\epsilon}{2}(q_{2}^{2}-q_{1}^{2})\right\} ,\label{3.7}
\end{gather}
which is consistent with equation~(\ref{25}). Using the same procedure we now wish to obtain the quantity $\langle p^{\ast}_{2},t_{2}|p_{1},t_{1}\rangle$. Again, we begin by calculating its variation
\begin{gather*}
\delta \langle p_{2}^{\ast},t_{2}|p_{1},t_{1}\rangle=\langle p_{2}^{\ast},t_{2}|G_{p_{2}^{\ast}}-G_{p_{1}}- H\delta T|p_{1},t_{1}\rangle ,
\end{gather*}
where now
\begin{gather*}
G_{p_{2}^{\ast}}=-q(t_{2})\delta p_{2}^{\ast} ,\qquad G_{p_{1}}=q(t_{1})\delta p_{1} .
\end{gather*}
We now need to f\/ix the initial and f\/inal conditions at $t_{1}$ and $t_{2}$ as $p(t_{1})$ and $p^{\dagger}(t_{2})$. Thus, again we get (\ref{solucionq}) and (\ref{solucionp}) as solutions to the Heisenberg equations, but now with coef\/f\/icients~$A$ and~$B$ given by
\begin{gather*}
A = \frac{1}{\left(\left(\epsilon^{2}+1\right)\sin(T)-2i\epsilon\cos(T)\right)} \\
\phantom{A=}{}\times \bigl\{ \bigl(\cos(t_{2})+i\epsilon \sin(t_{2})\bigr)p(t_{1})-\left(\cos(t_{1})-i\epsilon \sin(t_{1})\right)p^{\dagger}(t_{2})\bigr\}, \\
B = \frac{1}{\left(\left(\epsilon^{2}+1\right)\sin(T)-2i\epsilon\cos(T)\right)} \\
\phantom{B=}{}\times \bigr\{\bigl(\sin(t_{2})-i\epsilon \cos(t_{2})\bigl)p(t_{1})-\left(i\epsilon\cos(t_{1})+\sin(t_{1})\right)p^{\dagger}(t_{2})\bigr\}.
\end{gather*}
In this way, we can express $q(t_{1})$ and $q(t_{2})$ in terms of the initial and f\/inal conditions $p(t_{1})$ and~$p^{\dagger}(t_{2})$. The result is
\begin{gather*}
q(t_{1}) = \frac{1}{\left(\left(\epsilon^{2}+1\right)\sin(T)-2i\epsilon\cos(T)\right)}\bigl\{\bigl(\cos(T)+i\epsilon \sin(T)\bigr)p(t_{1})-p^{\dagger}(t_{2})\bigr\} ,\\
q(t_{2})= \frac{1}{\left((\epsilon^{2}+1) \sin(T)-2i\epsilon \cos(T)\right)}\bigl\{p(t_{1})-\left(\cos(T)+i\epsilon \sin(T)\right)p^{\dagger}(t_{2})\bigr\} .
\end{gather*}
One can also calculate the commutator
\begin{gather*}
\big[p(t_{1}),p^{\dagger}(t_{2})\big]=i\left(\left(\epsilon^{2}+1\right)\sin(T)-2i\epsilon\cos(T)\right) .
\end{gather*}
Using all of the above, we f\/ind the following expression for $H$ in terms of the operators $p^{\dagger}(t_{1})$ and $p(t_{0})$
\begin{gather*}
H = \frac{1}{2\left((\epsilon^{2}+1)\sin(T)-2i\epsilon\cos(T)\right)^{2}} \\
\phantom{H=}{}\times \Bigl\{(1-\epsilon^{2})(p^{2}(t_{1})+p^{\dagger\,2}(t_{2}))
-2\bigl((\epsilon^{2}+1)\cos(T)+2i\epsilon\sin(T)\bigr)p^{\dagger}(t_{2})p(t_{1})\Bigr\} \\
\phantom{H=}{}-\frac{i}{2}\frac{\left((\epsilon^{2}+1)\,\cos(T)+2i\epsilon\sin(T)\right)}
{\left((\epsilon^{2}+1)\sin(T)-2i\epsilon\cos(T)\right)} .
\end{gather*}
Since $p^{\dagger}(t_{2})$ and $p(t_{1})$ act according to equation (\ref{3.2}) we get
\begin{gather*}
 \delta \langle p_{2}^{\ast},t_{2}|p_{1},t_{1}\rangle=i\langle p_{2}^{\ast},t_{2}|\biggl\{\frac{1}{\left((\epsilon^{2}+1)\sin(T)-2i\epsilon\cos(T)\right)} \\
\qquad{}\times \Bigl\{\left((\cos(T)+i\epsilon\sin(T))p_{2}^{\ast}-p_{1}\right)\delta p_{2}^{\ast}+\left((\cos(T)+i\epsilon\sin(T))p_{1}-p_{2}^{\ast}\right)\delta p_{1}\Bigr\} \\
\qquad{} - \biggl(\frac{1}{2\left((\epsilon^{2}+1)\sin(T)-2i\epsilon\cos(T)\right)^{2}}
 \Bigl\{(1-\epsilon^{2})(p_{1}^{2}+p_{2}^{\ast\,2})-2((\epsilon^{2}+1)\cos(T) \\
\qquad{} +2i\epsilon\sin(T))p_{2}^{\ast}p_{1}\Bigr\}-\frac{i}{2}\frac{\left((\epsilon^{2}+1)\cos(T)
 +2i\epsilon\sin(T)\right)}{\left((\epsilon^{2}+1)\sin(T)+2i\epsilon\cos(T)\right)}\biggr)\delta T\biggr\}|p_{1},t_{1}\rangle
\\
\qquad{} =i\langle p_{2}^{\ast},t_{2}|\delta\Biggl\{\frac{1}{2\left((\epsilon^{2}+1)\sin(T)-2i\epsilon\cos(T)\right)}
\Bigl\{\left(\cos(T)+i\epsilon\sin(T)\right)(p_{2}^{\ast\,2}+p_{1}^{2})-2p_{2}^{\ast}p_{1}\Bigr\}\!\! \\
\qquad{} -i\ln\left(\frac{1}{\sqrt{(\epsilon^{2}+1)\sin(T)-2i\epsilon\cos(T)}}\right) \Biggr\}|p_{1},t_{1}\rangle .
\end{gather*}
Integration produces
\begin{gather*}
 \langle p_{2}^{\ast},t_{2}|p_{1},t_{1}\rangle=\frac{C}{\sqrt{(\epsilon^{2}+1)\sin(T)-2i\epsilon\cos(T)}} \\
\phantom{\langle p_{2}^{\ast},t_{2}|p_{1},t_{1}\rangle=}{}\times \exp\left\{\frac{i}{2}\frac{\left((\cos(T)+i\epsilon\sin(T))(p_{2}^{\ast\,2}+p_{1}^{2})-2p_{2}^{\ast}p_{1}\right)}{\left((\epsilon^{2}+1)\,\sin(T)-2i\epsilon\cos(T)\right)}
\right\} ,
\end{gather*}
where $C$ is a normalization constant. We note that
\begin{gather*}
\lim_{t_{2}\rightarrow t_{1}}\langle p_{2}^{\ast},t_{2}|p_{1},t_{1}\rangle=C\sqrt{-\frac{1}{2i\epsilon}}
\exp\left\{-\frac{1}{4\epsilon}(p_{1}-p_{2}^{\ast})^{2}\right\} .
\end{gather*}
Furthermore, in the limit $\epsilon\rightarrow0$, all eigenvalues $p$ cease to be complex, so in the double limit we must have
\begin{gather*}
\lim_{\epsilon\rightarrow0}\langle p_{2}^{\ast},t_{2}\rightarrow t_{1}|p_{1},t_{1}\rangle=\delta\left(p_{2}-p_{1}\right).
\end{gather*}
Fixing the constant $C$ accordingly we obtain
\begin{gather}
 \langle p_{2}^{\ast},t_{2}|p_{1},t_{1}\rangle=\frac{1}{\sqrt{2\pi i}}\frac{1}{\sqrt{(\epsilon^{2}+1)\sin(T)-2i\epsilon\cos(T)}} \nonumber\\
\phantom{\langle p_{2}^{\ast},t_{2}|p_{1},t_{1}\rangle=}{}\times
\exp\left\{\frac{i}{2}\frac{\left((\cos(T)+i\epsilon\sin(T))(p_{2}^{\ast\,2}
+p_{1}^{2})-2p_{2}^{\ast}p_{1}\right)}{\left((\epsilon^{2}+1) \sin(T)-2i\epsilon \cos(T)\right)}
\right\}. \label{49}
\end{gather}
Next we compare the standard harmonic oscillator propagator with that obtained in equation~(\ref{49}).
The bracket corresponding to the change of basis is readily shown to be
\begin{gather*}
\langle P|p \rangle = \sqrt{\frac{1}{2\pi\epsilon}} \mathrm{e}^{-\frac{1}{2\epsilon} \left( p-P \right)^2} ,
\end{gather*}                                                                                      which allow us to relate the propagator expressed in terms of Hermitian variables, $(\hat P,\hat Q)$, with that written in terms of non-Hermitian ones, $(\hat p,\hat q)$, as follows
\begin{gather*}
\langle p^{\ast},t_2| p',t_1\rangle = \frac{1}{2\pi\epsilon} \int dPdP' \mathrm{e}^{-\frac{1}{2}((p^{\ast}-P')^2+(p'-P)^2)}
\langle P',t_2|P,t_1 \rangle .
\end{gather*}
We have explicitly verif\/ied the above expression.
The completeness relation in the non-Hermitian description results
\begin{gather*}
\int d^2p\, \mu(p,p^*) |p\rangle\langle p^*| = 1 ,\qquad
\mu =\frac{1}{\sqrt{\pi\epsilon}}\mathrm{e}^{+\frac{1}{4\epsilon}(p-p^*)^2} .
\end{gather*}

\subsection{Path integral approach}\label{section3.3}

Consider the action for the modif\/ied harmonic oscillator in terms of the real and imaginary parts of $q$ and $p$,
\begin{gather*}
      S= \int _{t_1 }^{t_2 } {dt} \left(p\dot q - \left(\frac{p^2}{2} + \frac{1}{2}(1 - \epsilon^2 )q^2 + {i\epsilon} pq \right) \right) =  \int _{t_1 }^{t_2 } {dt} \Bigg( \left( {p_R + ip_I } \right)\left( {{\dot q}_R + i{\dot q}_I } \right) \\
       \phantom{S=}{}- \Bigg( \frac{{\left( {p_R + ip_I } \right)^2 }}{2} + \frac{1}{2}(1 - \epsilon^2 )\left( {q_R + iq_I } \right)^2 + {i\epsilon}  ( {p_R + ip_I } ) ( {q_R + iq_I } ) \Bigg) \Bigg).
  \end{gather*}
To compute the propagator using the path integral of this system we need to select an integration contour given by the reality conditions (\ref{realitycon}), that in terms of real and imaginary parts~is
\begin{gather*}
 q_I = 0, \qquad
 p_I = - \varepsilon q_R .
\end{gather*}
These conditions are a pair of second class constraints in the framework of the Dirac's method of quantization \cite{Henneaux,dirac}, so the path integral subjected to them is exactly the
Senjanovic path integral~\cite{Senjanovic}. In our case we obtain
\begin{gather*}
\left\langle {q_2 ,t_2 \left| {q_1 ,t_1 } \right.} \right\rangle = \int {Dp_R Dp_I Dq_R Dq_I \delta \left( {q_I } \right)} \delta \left( {p_I + \varepsilon q_R } \right)\exp \left( {iS} \right),
\end{gather*}
since the determinant of the Poisson bracket of the constraints is one. Integrating over the imaginary parts using the delta functionals results in
\begin{gather*}
\langle {q_2 ,t_2} | {q_1 ,t_1 }\rangle = \int {Dp_R Dq_R} \exp \left( i \int_{t_1 }^{t_2 } {dt} \left(\left(p_R - i\epsilon q_R \right)\dot q_R  - \left( \frac{p^2 _R}{2} + \frac{q_R^2}{2} \right) \right)\right),
\end{gather*}
Here we see that the path integral is reduced to the usual path integral of the harmonic oscillator plus a term that
contributes only to the classical action, so f\/inally the amplitude is
\begin{gather*}
\langle {q_2 ,t_2} | {q_1 ,t_1 }\rangle = \left(\frac{1}{2\pi i\hbar \sin T}\right)^{\frac{1}{2}}\exp\left( \frac{i}{2\sin T}\left[\left(q_1^2+q_2^2\right)\cos T-2 q_1 q_2\right]+\frac{\epsilon}{2}(q_2^2 -q_1^2) \right),
\end{gather*}
 with $T=t_2-t_1$. In this way we recover the result (\ref{3.7}).

 Summarizing, we have provided a complex canonical transformation taking the complex harmonic oscillator
 (\ref{Hmod}) into the ordinary harmonic oscillator so that the non-Hermitian variables fulf\/ill reality conditions
 in the Hilbert space ${\cal H}$ which are consistent with the canonical transformation. Let us observe that ${\cal PT}$ symmetry has not
 been invoked in this approach. The states~(\ref{psin}) possess positive def\/inite
 norm in $\cal H$ and a unitary time evolution follows from the canonical transformation.

\section[The modified PU oscillator as a second order theory]{The modif\/ied PU oscillator as a second order theory}\label{section4}

Now we proceed with the HOTD model.
Let us start with the PU oscillator Lagrangian
\begin{gather} \label{LPU}
L_{\mathrm{PU}} = -\frac{1}{2}\ddot{x}^2 + \frac{(\omega_1^2 + \omega_2^2)}{2} \dot{x}^2 - \frac{\omega_1^2 \omega_2^2}{2} x^2 ,
\end{gather}
where we assume $\omega_1>\omega_2$. $L_{\mathrm{PU}}$ is connected
to
\begin{gather}\label{Lxi}
L_{\xi} = \frac{1}{2} \dot{\xi_1}^2 - \frac{\omega_1^2}{2} \xi_1^2 + \frac{1}{2} \dot{\xi_2}^2 - \frac{\omega_2^2}{2} \xi_2^2 ,
\end{gather}
for real $\xi_i$, $i=1,2$, by the following relations
\begin{gather}
L_{\mathrm{PU}} + \frac{df}{dt} = L_{\xi}, \qquad f = \dot{x}\ddot{x} , \label{LL}\\
\xi_1 = i (ax+b\ddot{x}) , \qquad \xi_2 = cx + b\ddot{x} . \label{xi2}
\end{gather}
By choosing $a$, $b$, $c$ to be real we see that $x$ is necessarily complex. To accomplish (\ref{LL}), (\ref{xi2}) the following
values are obtained, for which the
same sign should be used,
\begin{gather} \label{coeffPU}
\frac{a}{\omega_2^2} = b = \frac{c}{\omega_1^2} = \pm \frac{1}{\sqrt{\omega_1^2-\omega_2^2}} .
\end{gather}
In other words, according to (\ref{LL}), $L_{\mathrm{PU}}$ in (\ref{LPU}) fails to be the real $L_{\xi}$ in~(\ref{Lxi}) only
by a total time derivative. Note that $f$ will give rise to the corresponding canonical transformation once it is expressed in terms
of phase space variables which we now derive. According to the Ostrogradsky method applied to~(\ref{LPU}) we get
\begin{gather*}
\Pi_x = \big(\omega_1^2 +\omega_2^2\big) \dot{x} + \dddot{x} ,\qquad
z = \dot{x} , \qquad
\Pi_z = - \ddot{x} ,
\end{gather*}
so that $\{x,\Pi_x\}=1$ and $\{z,\Pi_z\}=1$. Clearly $f=-z\Pi_z$. The quantum Hamiltonian is
\begin{gather}\label{HPU}
\hat H_{\mathrm{PU}} = - \frac{1}{2} \hat \Pi_z^2 - \frac{\omega_1^2
+ \omega_2^2}{2} \hat z^2 + \hat z \hat \Pi_x + \frac{\omega_1^2
\omega_2^2}{2} \hat x^2 .
\end{gather}
Using the afore mentioned transformation we have
\begin{alignat}{3}
& \hat x = ib\hat \xi_1 + b\hat \xi_2 , \qquad && \hat \Pi_x = ia \hat P_1 + c \hat P_2 , & \label{xxi1}\\
& \hat z = ib \hat P_1 + b \hat P_2 , \qquad && \hat \Pi_z = ic \hat \xi_1 + a \hat \xi_2 , & \label{xxi4}
\end{alignat}
where $\hat P_i$, $i=1,2$, are the canonical momenta conjugated to $\hat \xi_i$ obtained from (\ref{Lxi}). In terms of
these Hermitian variables the Hamiltonian (\ref{HPU}) takes the form
\begin{gather}\label{HHO}
\hat H_{\xi} = \frac{\hat P_1^2}{2} + \frac{\omega_1^2}{2} \hat \xi_1^2 +
\frac{\hat P_2^2}{2} + \frac{\omega_2^2}{2}\hat \xi_2^2 .
\end{gather}
So, starting from a Hamiltonian that is not bounded from below,
extending the canonical variables to the complex plane and adding a
total time derivative a well def\/ined Hamiltonian in terms of~$P_i$
and~$\xi_i$ is obtained. Notice that in~\cite{Bender:2007wu} it is
acknowledged there exists a similarity transformation relating the
original PU oscillator with a couple of independent harmonic
oscillators similar to our case. Nevertheless a further
transformation has to be supplied in~\cite{Bender:2007wu} to obtain
the right frequencies for the oscillators.

\subsection[The modified PU Lagrangian in the unequal frequency case in the extended phase space]{The modif\/ied PU Lagrangian in the unequal frequency case \\ in the extended phase space}\label{section4.1}

Instead of performing the canonical analysis starting from equation~(\ref{LPU}), we will consider (\ref{LL}) which includes the
additional time derivative. Namely we start from
\begin{gather*}
L_{\rm T}=\frac{1}{2}\ddot{x}^{2}+\frac{1}{2}\left( \omega
_{1}^{2}+\omega
_{2}^{2}\right) \dot{x}^{2}-\frac{1}{2}\omega _{1}^{2}\omega _{2}^{2}x^{2}+%
\dot{x}x^{(3)} .
\end{gather*}
We f\/irst notice that $L_{\rm T}=L_{\rm T}\left(
x,\dot{x},\ddot{x},x^{(3)}\right) $, so it is higher order than $L_{\text{PU}}$. This also means that, according to
Ostrogradsky's formalism, there will be a six-dimensional
phase space with coordinates $(x,\dot{x},\ddot{x})$ and momenta $%
(p_{0},p_{1},p_{2})$ given by
\begin{gather}
p_{0} = \left( \omega _{1}^{2}+\omega _{2}^{2}\right)
\dot{x}+x^{(3)} ,
\label{dp0} \\
p_{1} = 0 , \label{dp1} \\
p_{2} = \dot{x} , \label{dp2}
\end{gather}
according to the general def\/initions given in equation~(\ref{momen}). Now we notice that (\ref{dp1}) and~(\ref{dp2}) are in fact primary constraints, which we write as
\begin{gather}
\phi _{1} = p_{1}=0 , \label{dC1} \\
\phi _{2} = p_{2}-\dot{x}=0 . \label{dC2}
\end{gather}
The time evolution of the above constraints f\/ixes the corresponding
Lagrange multipliers making them second class constraints. In this
way, the reduced phase space has dimension four. We will choose to
express our Hamiltonian in terms of $(x,\ddot{x},p_{0},p_{2})$. The
resulting canonical Hamiltonian is given by
\begin{gather*}
H_{\rm C} = p_{0}\dot{x}+p_{1}\ddot{x}+p_{2}x^{(3)}-L_{\rm T}
 = p_{0}p_{2}-\frac{1}{2}\ddot{x}^{2}-\frac{1}{2}\big(\omega
_{1}^{2}+\omega _{2}^{2}\big)p_{2}^{2}+\frac{1}{2}\omega _{1}^{2}\omega
_{2}^{2}x^{2}.
\end{gather*}%
Next we construct the canonical structure in the Dirac formalism,
accounting for the second class constraints. To do so, we use the
Dirac brackets
\begin{gather*}
\left\{ A,B\right\} ^{\ast }=\left\{ A,B\right\} -\left\{ A,\phi
_{\alpha }\right\} C^{\alpha \beta }\left\{ \phi _{\beta },B\right\}
 ,\qquad \alpha ,\beta =1,2,
\end{gather*}
where $C^{\alpha \beta }$ is a square antisymmetric matrix such that
\begin{gather*}
C^{\mu \alpha }\left\{ \phi _{\alpha },\phi _{\nu }\right\} =\delta
_{\nu }^{\mu } .
\end{gather*}%
Upon using equations~(\ref{dC1}) and (\ref{dC2}) we obtain
\begin{gather*}
C^{\alpha \beta }=\left(
\begin{array}{cc}
0 & -1 \\
1 & 0%
\end{array}%
\right) .
\end{gather*}%
The resulting Dirac brackets are
\begin{gather}
\left\{ x,p_{0}\right\} ^{\ast }=1 ,\qquad \left\{
\ddot{x},p_{2}\right\} ^{\ast }=1 , \label{DB0}
\end{gather}
and zero for any other combination. Next we rewrite the Lagrangian
in the canonical formalism
\begin{gather*}
L_{\rm T}=\dot{x}p_{0}+\ddot{x}p_{1}+x^{(3)}p_{2}-H_{\rm C}
\end{gather*}
and make the constraints (\ref{dC1}), (\ref{dC2}) strong obtaining%
\begin{gather}
L_{\rm T}=\dot{x}p_{0}+x^{(3)}p_{2}-H_{\rm C}. \label{LAG0}
\end{gather}

In order to make contact with the classical version of the
Hamiltonian (\ref{HPU}) we introduce the following canonically
related new variables, leaving $x$ unchanged,
\begin{gather*}
 \Pi _{x}=p_{0} ,\qquad z=p_{2} ,\qquad \Pi
_{z}=-\ddot{x} . 
\end{gather*}%
Using equation~(\ref{DB0}), we obtain as the only non-zero Dirac brackets
\begin{gather*}
\left\{ x,\Pi _{x}\right\} ^{\ast }=1 ,\qquad \left\{ z,\Pi
_{z}\right\} ^{\ast }=1 .
\end{gather*}
In this way, we can consider the variables $\left\{ x,z,\Pi _{x},\Pi
_{z}\right\} $ as a set of canonical variables in the Dirac
formalism. Introducing the new variables in equation~(\ref{LAG0}) we
obtain
\begin{gather*}
L_{\rm T}=\dot{x}\Pi _{x}-\dot{\Pi}_{z}z-H_{\rm C}, 
\end{gather*}
where $H_{\rm C}$ corresponds to the classical counterpart of $H_{\rm PU}$
given in equation~(\ref{HPU}). The above equation guarantees that
$H_{\rm PU}$ is the correct Hamiltonian in the reduced physical space
and imply that the correct boundary conditions at the end points are
given by f\/ixing~$x$ and~$\Pi_{z}$ in regard to the variational
problem.

\subsection[The Pais-Uhlenbeck transformation in matrix form]{The Pais--Uhlenbeck transformation in matrix form}\label{section4.2}

The complex canonical transformation between the real variables
$\{\xi_{1},\xi_{2},P_{1},P_{2}\}$ and the complex ones
$\{x,z,\Pi_{x},\Pi_{z}\}$, is given in matrix form by
\begin{gather*}
X=M \xi ,\qquad \xi=M^{-1} X ,
\end{gather*}
where $X$, $\xi$ and $M$ stand for
\begin{gather*}
X=\left(\begin{array}{c}
x\\
z\\
\Pi_{x}\\
\Pi_{z}
\end{array}
\right) ,  \qquad
\xi=\left(\begin{array}{c}
\xi_{1}\\
\xi_{2}\\
P_{1}\\
P_{2}
\end{array}
\right) , \qquad M=\left(
\begin{array}{cccc}
ib & b & 0 & 0 \\
0 & 0 & ib & b\\
0& 0 & ia & c\\
ic& a& 0&0
\end{array}
\right) .
\end{gather*}
Now, the matrix $M$ satisf\/ies
\begin{equation}\label{simplec}
M^{T}\Omega M=\Omega ,
\end{equation}
with
\begin{gather*}
\Omega=
\left(
\begin{array}{cccc}
0 &0 & 1 & 0 \\
0 & 0 & 0 & 1\\
-1& 0 & 0 & 0\\
0& -1& 0&0
\end{array}
\right) ,
\end{gather*}
where only the relation $b(c-a)=1$ was needed to prove equation~(\ref{simplec}). This means that the matrix $M$ belongs to the complex symplectic group ${\rm Sp}\left(2n, \mathcal{C}\right)$, with $n=2$. Since ${\rm Sp}\left(2n, \mathcal{C}\right)$ is a~$2n(2n+1)$ parameter group, in our case $M$ should have $20$ parameters in general. One notices that
\begin{gather*}
\det (M)=(b(c-a))^2=1 ,
\end{gather*}
which verif\/ies the unimodular character of the symplectic matrix.

\subsection{Hilbert space}\label{section4.3}

We have related the Hamiltonians (\ref{HPU}) and (\ref{HHO}) by means of the complex canonical transformations (\ref{xxi1}), (\ref{xxi4}) and their inverse
\begin{alignat*}{3}
&\xi_{1}= iax-ib\Pi_{z}, \qquad &&
\xi_{2}= cx-b\Pi_{z},& \nonumber\\
&P_{\xi_{1}}=-icz+ib\Pi_{x}, \qquad &&
P_{\xi_{2}}=-az+b\Pi_{x},& 
\end{alignat*}
where the coef\/f\/icients $a$, $b$ and $c$ are given in (\ref{coeffPU}).
We notice that in this case the following relations are satisf\/ied:
\begin{gather*}
ac=\omega_{1}^{2}\omega_{2}^{2} b^{2}, \qquad b\left(c-a\right)=1 .
\end{gather*}

At the quantum mechanical level, we f\/ix the reality conditions by requiring that the operators $\left\{\xi_{1},\xi_{2},P_{\xi_{1}},P_{\xi_{2}}\right\}$ are Hermitian. This in turn implies that:
\begin{alignat}{3}
& x^{\dagger} = b\left(a+c\right)x-2b^{2}\Pi_{z}, \qquad &&  z^\dag= -b(a+c)z+ 2b^2 \Pi_x, & \label{xdagger}\\
& \Pi_{z}^{\dagger} = 2acx-b\left(a+c\right)\Pi_{z}, \qquad && \Pi_x^\dag = -2ac z +b(a+c) \Pi_x. & \label{pizdagger}
\end{alignat}
We will quantize the system in the basis $|x,\Pi_{z}\rangle$, so we propose the following operator realizations
\begin{gather*}
\hat{x}=x ,\qquad \hat{\Pi}_{z}=\Pi_{z} ,
\end{gather*}
and
\begin{gather*}
\hat{\Pi}_{x}=-i\frac{\partial}{\partial x} ,\qquad \hat{z}=i\frac{\partial}{\partial \Pi_{z}} .
\end{gather*}
Now, the inner product written in the basis $|x,\Pi_{z}\rangle$ can be expressed in general as
\begin{gather*}
\langle \phi,\psi\rangle=\int d^{2}xd^{2}\Pi_{z} \, \mu\left(x,\Pi_{z}\right)\phi^{\ast}\left(x,\Pi_{z}\right)\psi\left(x,\Pi_{z}\right),
\end{gather*}
where
\begin{gather*}
d^{2}x=dx_R dx_I ,\qquad d^{2}\Pi_{z}=d\Pi_{zR}d\Pi_{zI} .
\end{gather*}
Because of conditions (\ref{xdagger}) and (\ref{pizdagger}), we should have in particular:
\begin{gather}
\langle x \phi,\psi\rangle=\langle \phi,\left(b\left(a+c\right)x-2b^{2}\Pi_{z}\right)\psi\rangle ,\label{prodx}
\end{gather}
and
\begin{gather}
\langle \Pi_{z}\phi,\psi\rangle=\langle \phi,\left(2acx-b\left(a+c\right)\Pi_{z}\right)\psi\rangle .\label{prodpiz}
\end{gather}
or
\begin{gather*}
\int d^{2}xd^{2}\Pi_{z}\mu\, x^{\ast}\phi^{\ast}\psi=\int d^{2}xd^{2}\Pi_{z}\mu\,\phi^{\ast}\left\{b\left(a+c\right)x-2b^{2}\Pi_{z}\right\}\psi,\\
\int d^{2}xd^{2}\Pi_{z}\mu \, \Pi_{z}^{\ast}\phi^{\ast}\psi=\int d^{2}xd^{2}\Pi_{z}\mu\,\phi^{\ast}\left\{2acx-b\left(a+c\right)\Pi_{z}\right\}\psi.
\end{gather*}
Separating $x$ and $\Pi_{z}$ in their real and imaginary parts inside the integral,
and matching both equations, we conclude from (\ref{prodx}) and (\ref{prodpiz}) that:
\begin{gather}
a x_R = b \Pi_{zR} ,\label{condre}\\
c x_I = b \Pi_{zI} ,\label{condim}
\end{gather}
which can be always accomplished if
\begin{gather*}
\mu\left(x,\Pi_{z}\right)=C \delta\left(a x_R -b \Pi_{zR}\right)\delta\left(c x_I -b \Pi_{zI}\right) ,
\end{gather*}
where $C$ is a constant to be determined by some normalization condition.

\subsection{The Schr\"odinger equation}\label{section4.4}

Let us consider the stationary Sch\"odinger equation
\begin{gather*}
H_{M}\psi\left(x,\Pi_{z}\right)=E\,\psi\left(x,\Pi_{z}\right) ,
\end{gather*}
or more explicitly, using the operator realizations given above
\begin{gather*}
-\frac{1}{2}\Pi_{z}^{2}\psi+\frac{1}{2}\left(\omega_{1}^{2}+\omega_{2}^{2}\right)
\frac{\partial^{2}\psi}{\partial\Pi_{z}^{2}}+\frac{1}{2}\omega_{1}^{2}\omega_{2}^{2}x^{2}\psi+
\frac{\partial^{2}\psi}{\partial x\partial \Pi_{z}}=E \psi .
\end{gather*}
Using the relation between $\{\xi_{1},\xi_{2}\}$ and $\{x,\Pi_{z}\}$, applying the chain rule
\begin{gather*}
\frac{\partial}{\partial x} = \sum_{i}\frac{\partial \xi_{i}}{\partial x}\frac{\partial}{\partial \xi_{i}}=ia\frac{\partial}{\partial \xi_{1}}+c\frac{\partial}{\partial \xi_{2}} ,\\
\frac{\partial}{\partial \Pi_{z}} = \sum_{i}\frac{\partial \xi_{i}}{\partial \Pi_{z}}\frac{\partial}{\partial \xi_{i}}=-ib\frac{\partial}{\partial \xi_{1}}-b\frac{\partial}{\partial \xi_{2}} ,
\end{gather*}
and substituting we arrive at
\begin{gather*}
-\frac{1}{2}\frac{\partial^{2}\psi}{\partial \xi_{1}^{2}}-\frac{1}{2}\frac{\partial^{2}\psi}{\partial \xi_{2}^{2}}+\frac{\omega_{1}^{2}}{2}\xi_{1}^{2}\psi+\frac{\omega_{2}^{2}}{2}\xi_{2}^{2}\psi=E \psi ,
\end{gather*}
with regular solutions:
\begin{gather*}
\psi_{m,n}\left(\xi_{1},\xi_{2}\right)=\exp\left\{-\frac{1}{2}\left(\omega_{1}\xi_{1}^{2}
+\omega_{2}^{2}\xi_{2}^{2}\right)\right\}H_{m}\left(\sqrt{\omega_{1}}\xi_{1}\right)H_{n}\left(\sqrt{\omega_{2}}\xi_{2}\right) .
\end{gather*}
Let us now consider the ground state, $m=n=0$
\begin{gather*}
\psi_{0,0}\left(\xi_{1},\xi_{2}\right)=\exp\left\{-\frac{1}{2}\left(\omega_{1}\xi_{1}^{2}+\omega_{2}^{2}\xi_{2}^{2}\right)\right\} ,
\end{gather*}
or written in terms of $x$, $\Pi_{z}$
\begin{gather*}
\psi_{0,0}\left(x,\Pi_{z}\right)=\exp\left\{\frac{1}{2}\left(\left(\omega_{1}a^{2}
-\omega_{2}c^{2}\right)x^{2}+2b\left(\omega_{2}c-\omega_{1}a\right)\Pi_{z}x+b^{2}\left(\omega_{1}-\omega_{2}\right)\Pi_{z}^{2}\right)\right\} .
\end{gather*}
Separating $x$ and $\Pi_{z}$ in their real and imaginary parts we get
\begin{gather}
\psi_{0,0}\left(x,\Pi_{z}\right) = \exp\biggl\{\frac{1}{2}\Bigl(\left(\omega_{1}a^{2}-
\omega_{2}c^{2}\right)\left(x_R^2-x_I^2\right)
 +2b\left(\omega_{2}c-\omega_{1}a\right)\left(\Pi_{zR} x_R -\Pi_{zI} x_I\right)\nonumber\\
\phantom{\psi_{0,0}\left(x,\Pi_{z}\right) =}{} +b^{2}\left(\omega_{1}-\omega_{2}\right)\left(\Pi_{zR}^2 -\Pi_{zI}^2\right)\Bigr)\biggr\}\exp\left\{i \phi\left(x,\Pi_{z}\right)\right\} ,\label{ground}
\end{gather}
where $\phi\left(x,\Pi_{z}\right)$ is a real function of $x_R$, $x_I$, $\Pi_{zR}$ and $\Pi_{zI}$ and therefore is the imaginary part of $\psi_{0,0}$ written as a phase factor. It is not easy to see from (\ref{ground}) if the ground state function vanishes at $|x|\rightarrow \pm \infty$, $|\Pi_{z}|\rightarrow \pm \infty$. However, substituting the reality conditions (\ref{condre}), (\ref{condim}) we have
\begin{gather*}
\psi_{0,0}\left(z,\Pi_{z}\right)=\exp\left\{-\frac{\omega_{2}\left(\omega_{1}^{2}-
\omega_{2}^{2}\right)}{2} x_R^2\right\}\exp\left\{-\frac{\omega_{1}\left(\omega_{1}^{2}-\omega_{2}^{2}\right)}{2}x_I^2\right\} ,
\end{gather*}
where we verify that the ground state function indeed vanishes at inf\/inity.

Finally, in the Appendix we compute the full propagator of the PU oscillator.

\subsection[The modified PU Lagrangian in the equal frequency case as a third order system]{The modif\/ied PU Lagrangian in the equal frequency case\\ as a third order system}\label{section4.5}

We now wish to follow the same procedure as in Subsection~\ref{section4.1} for
the restriction of the PU Lagrangian to the equal frequency case.
Recapitulating, we begin from the corresponding equal frequency PU
Lagrangian $\tilde{L}_{\rm PU}$ plus a total time derivative
\begin{gather}
\tilde{L}_{\rm T}=\tilde{L}_{\rm PU}+\frac{d\tilde{f}}{dt}=-\frac{1}{2}\ddot{x}%
^{2}+\omega ^{2}\dot{x}^{2}-\frac{1}{2}\omega ^{4}x^{2} +\frac{d\tilde{f}}{%
dt}. \label{MODEFPU}
\end{gather}
The equations of motion arising from $\tilde{L}_{\rm PU}$ are%
\begin{gather*}
0=x^{(4)}+2\omega ^{2}x^{(2)}+\omega ^{4}x, 
\end{gather*}
which are not af\/fected by the total time derivative.

At the Lagrangian level the choice
\begin{gather*}
\tilde{f}=\frac{1}{2}\left( \ddot{x}-\omega ^{2}x\right) \dot{x}
\end{gather*}
allows to transform (\ref{MODEFPU}) into the two oscillators system
\begin{gather*}
\tilde{L}_{\rm PU}+\frac{d\tilde{f}}{dt}=\tilde{L}_{\xi }=\frac{1}{2}\dot{\xi}%
_{1}^{2}+\frac{1}{2}\dot{\xi}_{2}^{2}-\frac{1}{2}\Omega ^{2}\left(
\xi _{1}^{2}+\xi _{2}^{2}\right) , 
\end{gather*}
by means of the nonlocal transformation
\begin{gather*}
\xi _{1} = i\left( ax+b\ddot{x}\right) , \qquad 
\xi _{2} = cx+d\ddot{x} . 
\end{gather*}%
together with the conditions
\begin{gather*}
\omega ^{2} = \Omega ^{2} ,\qquad b^{2}=d^{2} , \qquad
cd-ab = 1/2,\qquad c^{2}-a^{2}=\omega ^{2}.
\end{gather*}
For simplicity we can take $d=b$, which gives us
\begin{gather*}
a=\omega ^{2}b-\frac{1}{4b} ,\qquad c=\omega ^{2}b+\frac{1}{4b} .
\end{gather*}
Then, in this case we have a free parameter in the transformation.
Now, the total Lagrangian turns out to be
\begin{gather*}
\tilde{L}_{\rm T}=\frac{1}{2}\omega ^{2}\dot{x}^{2}-\frac{1}{2}\omega ^{4}x^{2}+%
\frac{1}{2}x^{(3)}\dot{x}-\frac{1}{2}\omega ^{2}x\ddot{x} ,
\end{gather*}%
where $x^{(3)}=\dddot{x}$. According to Ostrogradsky, the canonical
momenta conjugated to the coordinates $(x,\dot{x},\ddot{x})$ are
\begin{gather}
\tilde{p}_{0} = \frac{3}{2}\omega ^{2}\dot{x}+x^{(3)} , \label{p0b} \\
\tilde{p}_{1} = -\frac{1}{2}\left( \omega ^{2}x+\ddot{x}\right) ,\label{p1b} \\
\tilde{p}_{2} = \frac{1}{2}\dot{x}, \label{p2b}
\end{gather}
respectively. Again, (\ref{p1b}) and (\ref{p2b}) are primary
constraints, which can be put as
\begin{gather}
\phi _{1}=\tilde{p}_{1}+\frac{1}{2}(\omega
^{2}x+\ddot{x})=0 ,\qquad \phi
_{2}=\tilde{p}_{2}-\frac{1}{2}\dot{x}=0. \label{PCEQFEC}
\end{gather}
As in the previous case, the time evolution of them f\/ixes the
corresponding Lagrange multipliers, thus turning (\ref{PCEQFEC})
into a set of second class constraints which reduce the dimension of
the phase space from six to
four. We choose to express our Hamiltonian in terms of variables $\left\{ x,%
\ddot{x},\tilde{p}_{0},\tilde{p}_{2}\right\} $, eliminating $\dot{x}$, $\tilde{p}_{1}$ from the constraints. The result is
\begin{gather*}
\tilde{H}_{\rm C} = \tilde{p}_{0}\dot{x}+\tilde{p}_{1}\ddot{x}+\tilde{p}%
_{2}x^{(3)}-\tilde{L}_{\rm T}  
 =2\tilde{p}_{2}(\tilde{p}_{0}+\omega _{2}\tilde{p}_{2})-4\omega
^{2}\tilde{p}_{2}^{2}+\frac{1}{2}\omega
^{4}x^{2}-\frac{1}{2}\ddot{x}^{2} . 
\end{gather*}%
A similar procedure as the one described in Subsection~\ref{section4.1} yields
the non-zero Dirac brackets
\begin{gather*}
\left\{ x,\tilde{p}_{0}\right\} ^{\ast }=1 ,\qquad \left\{ \ddot{x},\tilde{%
p}_{0}\right\} ^{\ast }=-\frac{1}{2}\omega ^{2} ,\qquad \left\{ \ddot{x},%
\tilde{p}_{2}\right\} ^{\ast }=\frac{1}{2} , 
\end{gather*}
in the phase space $\left\{
x,\ddot{x},\tilde{p}_{0},\tilde{p}_{2}\right\} .$ In order to make
contact with the original equal frequency PU problem, let us
introduce the new variables $\tilde{x}$, $\tilde{z}$,
$\tilde{\Pi}_{x}$, $\tilde{\Pi}_{z}$
\begin{equation*}
\tilde{x}=x,\qquad \tilde{z}=2\tilde{p}_{2} ,\qquad \tilde{\Pi}_{x}=\tilde{p}_{0}+\omega ^{2}\tilde{p}_{2},\qquad \tilde{\Pi}_{z}=-\ddot{x},
\end{equation*}
and proceed to calculate their Dirac brackets. The only non-zero
results in the phase space
$\{\tilde{x},\tilde{z},\tilde{\Pi}_{x},\tilde{\Pi}_{z}\}$ are
\begin{gather*}
\big\{ \tilde{x},\tilde{\Pi}_{x}\big\} ^{\ast }=1,\qquad \big\{ \tilde{z},\tilde{\Pi}_{z}\big\} ^{\ast }=1,
\end{gather*}
The above indicates that we can consider the set of variables $\{\tilde{x},
\tilde{z},\tilde{\Pi}_{x},\tilde{\Pi}_{z}\}$ as canonical variables
in the Dirac formalism. Substituting in ${\tilde H}_{\rm C}$ we get:
\begin{gather}
\tilde{H}_{\rm C}=-\frac{1}{2}\tilde{\Pi}_{z}^{2}-\omega ^{2}\tilde{z}^{2}+%
\tilde{\Pi}_{x}\tilde{z}+\frac{1}{2}\omega ^{4}\tilde{x}^{2},
\label{HPUEQFREC}
\end{gather}%
which corresponds to the original Pais--Uhlenbeck Hamiltonian~\eqref{HPU}
in the equal-frequency case. As pointed out in references
\cite{Bender:2008vh1,Smilga1}, (\ref{HPUEQFREC}) is not
diagonalizable and hence cannot be related by a similarity
transformation to the two oscillator system.

However, the Pais--Uhlenbeck Hamiltonian in
the equal frequencies case can be related to the Hamiltonian of two
uncoupled oscillators
\begin{gather*}
H=\frac{1}{2}\left(P_{1}^{2}+P_{2}^{2}\right)+\frac{1}{2}\omega^{2}\left(\xi_{1}^{2}+\xi_{2}^{2}\right),
\end{gather*}
through a transformation which is neither canonical nor of the
similarity type, given by
\begin{gather*}
\left(
\begin{array}{c}
x \\
z  \\
\Pi_{x}\\
\Pi_{z}
\end{array}
\right)=\left(
\begin{array}{cccc}
ib & \sqrt{\frac{1}{\omega^{2}}+b^{2}}&0&0 \\
0 & 0& id&d\\
0&0& i\left(d\omega^{2}-\frac{1}{2d}\right)&\left(d\omega^{2}+\frac{1}{2d}\right)\\
i\omega^{2}\sqrt{\frac{1}{\omega^{2}}+b^{2}}&b\omega^{2}&0&0
\end{array}
\right)\left(
\begin{array}{c}
\xi_{1} \\
\xi_{2}  \\
P_{1}\\
P_{2}
\end{array}
\right),
\end{gather*}
or, conversely
\begin{gather*}
\left(
\begin{array}{c}
\xi_{1} \\
\xi_{2}  \\
P_{1}\\
P_{2}
\end{array}
\right)=\left(
\begin{array}{cccc}
ib\omega^{2} & 0&0&-i\sqrt{\frac{1}{\omega^{2}}+b^{2}} \\
\omega^{2}\sqrt{\frac{1}{\omega^{2}}+b^{2}} & 0& 0&-b\\
0&-i\left(d\omega^{2}+\frac{1}{2d}\right)& id&0\\
0&-\left(d\omega^{2}-\frac{1}{2d}\right)&d&0
\end{array}
\right)\left(
\begin{array}{c}
x \\
z  \\
\Pi_{x}\\
\Pi_{z}
\end{array}
\right).
\end{gather*}
 Even though we have not explored the quantization of the system in this
case, we expect that the non existence of a similarity
transformation between the PU oscillator and the two-oscillator
system would be a signal of the non-diagonalizable property of the
Hamiltonian related to the appearance of Jordan blocks, as discussed
in~\cite{Mannheim1,Bender:2008vh1,Smilga1}.

\section{Discussion}\label{section5}

In this work we have proposed that a complex canonical
transformation applied to a class of HOTD models with Lagrangians
modif\/ied by a total time derivative can solve the drawbacks of such
theories, namely negative norm states or ghosts, unbounded below
Hamiltonian and non-unitary time evolution, keeping unmodif\/ied the
original classical equations of motion. Such a~transformation
requires the original canonical variables to be analytically
continued and further subjected to appropriate reality conditions so
that the f\/inal canonical set matches the original number of real
degrees of freedom. In particular while the f\/inal canonical
variables~$\xi^i$ are Hermitian the original ones $x^{(i)}$ fulf\/ill
some reality conditions dictated by the canonical transformation
among such variables. The inner product in the original Hilbert
space is def\/ined in such way to reproduce the reality conditions.
This idea has been applied previously to non-perturbative canonical
quantum general relativity \cite{Ashtekar:1991hf,Ashtekar:1992ne}
and the inclusion of reality conditions was mandatory to recover
real gravity. Thus we have successfully provided a novel extension
of such idea to the case of the PU oscillator as an example of HOTD
models. An important point that we must clarify is that the analytic
extension of the variables allow us to have a well def\/ined
transformation between the HOTD model and a second order theory. The
interesting result is that both theories have the same degrees of
freedom and are related by a complex transformation. So, our
procedure to extend the original problem to the complex plane gives
us the appropriate frame to f\/ind the mapping between both theories.
The success of the proposed approach in the PU case motivates the
study of its extension to other HOTD models, but this is beyond the
scope of the present work.

To illustrate the procedure we have f\/irst reviewed the case of a
complexif\/ied model that reduces to an harmonic oscillator. Although
not of the HOTD type, this example exhibits negative norm states
which are neatly eliminated using a complex canonical
transformation. In this case we determined also the corresponding
propagator and contrasted it with the usual one of the harmonic
oscillator. To do so we adopted two dif\/ferent techniques:
Schwinger's quantum action principle and the path integral approach,
both of which led to the same result. Let us notice also that the
path integral version of complex canonical variables for gravity has
been studied \cite{Alexandrov:1998cu}.

Next we studied the PU oscillator (\ref{LPU}) which is properly a
HOTD model. After complexi\-fying the original variables we
constructed the complex canonical transformation
(\ref{xxi1}), (\ref{xxi4}) which leads to a system of two decoupled
harmonic oscillators (\ref{HHO}) with just the frequencies
$\omega_1$ and $\omega_2$ appearing in (\ref{LPU}). The propagator
of the model containing non-Hermitian variables was determined by
adopting again Schwinger's quantum action principle. The equal
frequency PU oscillator has been also described in the classical
case. We show that there are some dif\/ferences with the unequal
frequency situation. First, we have here a free parameter in the
transformation, perhaps a resemblance of the Barbero--Immirzi parameter
in loop quantum gravity \cite{Rovelli:2004tv,Thiemann:2007zz}.
Furthermore, in the reduced phase space there does not exist a
similarity transformation between the PU oscillator with equal
frequency and the second order oscillators. However, we found that
this system is also mapped to a couple of harmonic oscillators,
using a non-similarity type transformation.

Our approach to the PU oscillator is an alternative to that of
\cite{Bender:2007wu} based on non-Hermitian but ${\cal PT}$
symmetric Hamiltonians. We f\/ind ours has the following properties:
(1)~It is based on non-Hermitian variables subjected to specif\/ic and
well def\/ined reality conditions arising from
(\ref{xxi1}), (\ref{xxi4}) that make the higher-order Hamiltonian
Hermitian in the appropriate Hilbert space with measure
(\ref{meas}). (2)~No $\cal PT$ symmetry is required. (3)~It works
for at least the PU oscillator. (4)~If we want to consider
additional interaction terms as an anharmonic contribution, our
proposal still works provided that we add the prescription $x^4 \to
(xx^\dagger)^2=b^4(\xi_1^2 +\xi_2^2)^2$. This interaction term has
the correct signs and the Hamiltonian is still bounded from below.
(5)~We must emphasize that the complexif\/ication of the original
problem together with the reality conditions are a key ingredient in
our approach which constitute the prescription that def\/ine our
problem.

It is remarkable that this approach can be also applied to the case
of a quadratic higher-order derivative free scalar f\/ield model. To
see this notice that the following Lagrangian density
\begin{gather*}
{\cal L} = -\frac{1}{2} (\Box \phi)^2 - \frac{m_1^2+m_2^2}{2} \phi\Box\phi -\frac{m_1^2 m_2^2}{2} \phi^2 ,
\end{gather*}
is related to
\begin{gather}\label{Lpsi}
{\cal L}_{\psi} = -\frac{1}{2} \psi_1 \Box \psi_1 -\frac{m_1^2}{2} \psi_1^2 -\frac{1}{2} \psi_2 \Box \psi_2
-\frac{m_2^2}{2} \psi_2^2 ,
\end{gather}
by
\begin{alignat*}{3}
& {\cal L} + \partial_{\mu} f^{\mu} = {\cal L}_{\psi} ,\qquad &&
f^{\mu} = - \Box \phi \partial^{\mu}\phi  , & \\
& \psi_1 = i(a\phi + b\Box \phi) , \qquad && \psi_2 = c\phi + b\Box \phi ,&
\end{alignat*}
with $a$, $b$, $c$ having the same form as in the PU model (\ref{coeffPU})
except for the replacing $\omega_i\rightarrow m_i$, $i=1,2$. Clearly
(\ref{Lpsi}) is ghost-free. Our approach seems promising even in the
interacting case including a term of the type $(\phi\phi^{\ast})^2=
-\frac{1}{(c-a)^4}(\psi_1^2+\psi_2^2)^2$
\cite{Hawking:2001yt,Antoniadis:2006pc}, however all these
generalizations deserve further study.

In closing we would like to mention some directions for further
study along the lines proposed in the present work. Other physical
mechanical models including at least third order derivatives might
help to show the strength of the canonical approach here pursued.
Also a crucial problem to come to terms with is f\/ield theory. The
success in the simplest non interacting scalar f\/ield brief\/ly
described above requires full implementation at the quantum level.
Finally, as it was mentioned in the introduction, higher order
gravity has been one of the motivations to the present approach,
which of course is a paramount challenge. Also it would be rather
interesting to look for HOTD gauge f\/ield theories to test some
aspects of the gravitational case.

\appendix

\section{Schwinger's quantum action principle and the PU oscillator}\label{appendixA}

In this appendix we compute the propagator of the PU oscillator, a
result already obtained in~\cite{Mannheim}. To do this, we
apply Schwinger's quantum action principle to calculate $\langle
x_{2}^{\ast},\Pi_{z\,2}^{\ast},t_{2}|x_{1}$, $\Pi_{z\,1},t_{1}\rangle$.
However, in this case we need to be more careful since our variables
are complex. We will work in the Heisenberg picture, where the
involved operators act as:
\begin{alignat*}{3}
& x(t_{1})|x_{1},\Pi_{z 1},t_{1}\rangle=x_{1}|x_{1},\Pi_{z 1},t_{1}\rangle ,\qquad && \Pi_{z}(t_{1})|x_{1},\Pi_{z 1},t_{1}\rangle=\Pi_{z 1}|x_{1},\Pi_{z 1},t_{1}\rangle, &
\\
& \langle x_{2}^{\ast},\Pi_{z 2}^{\ast},t_{2}|x^{\dagger}(t_{2})=\langle x_{2}^{\ast},\Pi_{z 2}^{\ast},t_{2}|x_{2}^{\ast} ,\qquad && \langle x_{2}^{\ast},\Pi_{z 2}^{\ast},t_{2}|\Pi_{z}^{\dagger}(t_{2})=\langle x_{2}^{\ast},\Pi_{z 2}'^{\ast},t_{2}|\Pi_{z\,2}^{\ast} .&
\end{alignat*}
As usual, we f\/irst calculate the total variation of the quantity $\langle x_{2}^{\ast},\Pi_{z 2}^{\ast},t_{2}|x_{1},\Pi_{z 1},t_{1}\rangle$, given by:
\begin{gather}
 \delta\langle x_{2}^{\ast},\Pi_{z 2}^{\ast},t|x_{1},\Pi_{z 1}'',t_{1}\rangle\label{totvar}\\
 =i\langle x_{2}^{\ast},\Pi_{z 2}'^{\ast},t_{2}|\big\{\Pi_{x}^{\dagger}(t_{2})\delta x_{2}^{\ast}-z^{\dagger}(t_{2})\delta \Pi_{z 2}^{\ast}-\Pi_{x}(t_{1})\delta x_{1}+z(t_{1}) \delta \Pi_{z 1}-H\delta T\big\}|x_{1},\Pi_{z 1},t_{1}\rangle ,\nonumber
\end{gather}
where $T=t_{2}-t_{1}$. Now, we must express $\{z,\Pi_{x},z^{\dagger}(t),\Pi_{x}^{\dagger}(t)\}$ in terms of $\{x,\Pi_{z},x^{\dagger}(t),\Pi_{z}^{\dagger}(t)\}$.

We solve the Heisenberg equations for the operators $x$, $z$, $\Pi_{x}$ and $\Pi_{z}$. This equations are:
\begin{gather*}
\dot{x}(t) = z(t) ,\\
\dot{\Pi}_{x}(t) = -\omega_{1}^{2}\omega_{2}^{2}x(t) ,\\
\dot{z}(t) = -\Pi_{z}(t),\\
\dot{\Pi}_{z}(t) = \left(\omega_{1}^{2}+\omega_{2}^{2}\right) z(t)-\Pi_{x}(t) .
\end{gather*}

Uncoupling, we get the following equation for $x(t)$:
\[
x^{(4)}(t)+\left(\omega_{1}^{2}+\omega_{2}^{2}\right)\ddot{x}(t)+\omega_{1}^{2}\omega_{2}^{2}x(t)=0 ,
\]
with general solution:
\[
x(t)=C_{1}e^{i \omega_{1} t}+C_{2}e^{-i \omega_{1} t}+C_{3}e^{i \omega_{2} t}+C_{4}e^{-i \omega_{2} t} ,
\]
while for the other operators we get:
\begin{gather*}
z(t) = i \omega_{1}\left(C_{1}e^{i \omega_{1} t}-C_{2}e^{-i \omega_{1} t}\right)+i \omega_{2}\left(C_{3}e^{i \omega_{2} t}-C_{4}e^{-i \omega_{2} t}\right) ,\\
\Pi_{x}(t) = i \omega_{1}\omega_{2}^{2}\left(C_{1}e^{i \omega_{1} t}-C_{2}e^{-i \omega_{1} t}\right)+i \omega_{1}^{2}\omega_{2}\left(C_{3}e^{i \omega_{2} t}-C_{4} e^{-i \omega_{2} t}\right) ,\\
\Pi_{z}(t) = \omega_{1}^{2}\left(C_{1}e^{i \omega_{1} t}+C_{2}e^{-i \omega_{1} t}\right)+\omega_{2}^{2}\left(C_{3}e^{i \omega_{2} t}-C_{4}e^{-i \omega_{2} t}\right) .
\end{gather*}
We f\/ix the constants $C_{i}$'s in terms of initial conditions at time $t_{1}$: $x(t_{1})$, $\Pi_{z}(t_{1})$ and f\/inal conditions at time $t_{2}$: $x^{\dagger}(t_{2})$ and $\Pi_{z}^{\dagger}(t_{2})$. To obtain such conditions, we use the reality properties (\ref{xdagger}), (\ref{pizdagger}), valid at all times. In this way, we arrive to the following system of linear equations:
\begin{gather*}
x(t_{1}) = C_{1}e^{i\omega_{1}t_{1}}+C_{2}e^{-i\omega_{1}t_{1}}+C_{3}e^{i\omega_{2}t_{1}}+C_{4}e^{-i\omega_{2}t_{1}} , \\
\Pi_{z}(t_{1}) = \omega_{1}^{2}\left(C_{1}e^{i\omega_{1}t_{1}}+C_{2}e^{-i\omega_{1}t_{1}}\right)+ \omega_{2}^{2}\left(C_{3}e^{i\omega_{2}t_{1}}+C_{4}e^{-i\omega_{2}t_{1}}\right) , \\
x^{\dagger}(t_{2}) = -\left(C_{1}e^{i \omega_{1} t_{2}}+C_{2}e^{-i \omega_{1} t_{2}}\right)+\left(C_{3}e^{i \omega_{2} t_{2}}+C_{4}e^{-i \omega_{2} t_{2}}\right) , \\
\Pi_{z}^{\dagger}(t_{2}) = -\omega_{1}^{2}\left(C_{1}e^{i \omega_{1} t_{2}}+C_{2}e^{-i \omega_{1} t_{2}}\right)+\omega_{2}^{2}\left(C_{2}e^{i \omega_{2}t_{2}}+C_{4}e^{-\omega_{2} t_{2}}\right) ,
\end{gather*}
Solving for the $C$'s and substituting, we get the following expressions for $z(t_{1})$, $\Pi_{x}(t_{1})$, $z^{\dagger}(t_{2})$ and $\Pi_{x}^{\dagger}(t_{2})$:
\begin{gather*}
z(t_{1}) = \frac{1}{(\omega_{1}^{2}-\omega^{2}_{2})\sin(\omega_{1}T)\sin(\omega_{2}T)} \\
\phantom{z(t_{1}) =}{} \times \{\{-\omega_{1}^{2}\omega_{2}\sin(\omega_{1}T)\cos(\omega_{2}T)
+\omega_{1}\omega_{2}^{2}\cos(\omega_{1}T)\sin(\omega_{2}T)\}x(t_{1}) \\
\phantom{z(t_{1}) =}{} +\{\omega_{2}\sin(\omega_{1}T)\cos(\omega_{2}T)-\omega_{1}\cos(\omega_{1}T)\sin(\omega_{2}T)\}\Pi_{z}(t_{1}) \\
\phantom{z(t_{1}) =}{} +\{\omega_{1}^{2}\omega_{2}\sin(\omega_{1}T)+\omega_{1}\omega_{2}^{2}\sin(\omega_{2}T)\}
x^{\dagger}(t_{2})+\{-\omega_{2}\sin(\omega_{1}T)-\omega_{1}\sin(\omega_{2}T)\}\Pi_{z}^{\dagger}(t_{2})\},\!\\
\Pi_{x}(t_{1}) = \frac{1}{(\omega_{1}^{2}-\omega_{2}^{2})\sin(\omega_{1}T)\sin(\omega_{2}T)} \\
\phantom{\Pi_{x}(t_{1}) =}{}\times \{\{-\omega_{1}^{4}\omega_{2}\sin(\omega_{1}T)\cos(\omega_{2}T)
+\omega_{1}\omega_{2}^{4}\cos(\omega_{1}T)\sin(\omega_{2}T)\}x(t_{1}) \\
\phantom{\Pi_{x}(t_{1}) =}{} +\{\omega_{1}^{2}\omega_{2}\sin(\omega_{1}T)\cos(\omega_{2}T)
-\omega_{1}\omega_{2}^{2}\cos(\omega_{1}T)\sin(\omega_{2}T)\}\Pi_{z}(t_{1})
\\
\phantom{\Pi_{x}(t_{1}) =}{} +\{\omega_{1}^{4}\omega_{2}\sin(\omega_{1}T)+\omega_{1}\omega_{2}^{4}\sin(\omega_{2}T)\}
x^{\dagger}(t_{2})\\
\phantom{\Pi_{x}(t_{1}) =}{}
+\{-\omega_{1}^{2}\omega_{2}\sin(\omega_{1}t_{2})
-\omega_{1}\omega_{2}^{2}\sin(\omega_{2}T)\}\Pi_{z}^{\dagger}(t_{2})\},\\
z^{\dagger}(t_{2}) = \frac{1}{(\omega_{1}^{2}-\omega_{2}^{2})\sin(\omega_{1}T)\sin(\omega_{2}T)}
\{\{-\omega_{1}^{2}\omega_{2}\sin(\omega_{1}T)-\omega_{1}\omega_{2}^{2}\sin(\omega_{2}T)\}x(t_{1})
\\
\phantom{z^{\dagger}(t_{2}) =}{}
 +\{\omega_{2}\sin(\omega_{1}T)+\omega_{1}\sin(\omega_{2}T)\}\Pi_{z}(t_{1}) \\
\phantom{z^{\dagger}(t_{2}) =}{}  +\{\omega_{1}^{2}\omega_{2}\sin(\omega_{1}T)\cos(\omega_{2}T)-\omega_{1}\omega_{2}^{2}
 \cos(\omega_{1}T)\sin(\omega_{2}T)\}x^{\dagger}(t_{2}) \\
\phantom{z^{\dagger}(t_{2}) =}{} +\{-\omega_{2}\sin(\omega_{1}T)\cos(\omega_{2}T)+\omega_{1}\cos(\omega_{1}T)
\sin(\omega_{2}T)\}\Pi_{z}^{\dagger}(t_{2})\} ,
\\
\Pi_{z}^{\dagger}(t_{2}) = \frac{1}{(\omega_{1}^{2}-\omega_{2}^{2})\sin(\omega_{1}T)\sin(\omega_{2}T)}
\{\{-\omega_{1}^{4}\omega_{2}\sin(\omega_{1}T)-\omega_{1}\omega_{2}^{4}\sin(\omega_{2}T)\}x(t_{1})
\\
\phantom{\Pi_{z}^{\dagger}(t_{2}) =}{}  +\{\omega_{1}^{2}\omega_{2}\sin(\omega_{1}T)+\omega_{1}\omega_{2}^{2}\sin(\omega_{2}T)\}\Pi_{z}(t_{1}) \\
\phantom{\Pi_{z}^{\dagger}(t_{2}) =}{} +\{\omega_{1}^{4}\omega_{2}\sin(\omega_{1}T)\cos(\omega_{2}T)
-\omega_{1}\omega_{2}^{4}\cos(\omega_{1}T)\sin(\omega_{2}T)\}x^{\dagger}(t_{2}) \\
\phantom{\Pi_{z}^{\dagger}(t_{2}) =}{} +\{-\omega_{1}^{2}\omega_{2}\sin(\omega_{1}T)\cos(\omega_{2}T)
+\omega_{1}\omega_{2}^{2}\cos(\omega_{1}T)\sin(\omega_{2}T)\}\Pi_{z}^{\dagger}(t_{2})\} .
\end{gather*}
Using these results in the Hamiltonian (\ref{HPU}) and in the total variation (\ref{totvar}), we get f\/inally
\begin{gather*}
 \langle x_{2}^{\ast},\Pi_{z\,2}^{\ast},t_{2}|x_{1},\Pi_{z\,1},t_{1}\rangle=
\sqrt{\frac{1}{\sin(\omega_{1}T)\sin(\omega_{2}T)}} \exp\Biggl\{\frac{i}{(\omega_{1}^{2}-\omega_{2}^{2})\sin(\omega_{1}T)\sin(\omega_{2}T)} \\
\qquad{} \times\Bigg\{ \{\omega_{1}^{4}\omega_{2}\sin(\omega_{1}T)\cos(\omega_{2}T)
-\omega_{1}\omega_{2}^{4}\cos(\omega_{1}T)\sin(\omega_{2}T)\}
\left\{\frac{x_{1}^{2}}{2}+\frac{x_{2}^{\ast\,2}}{2}\right\} \\
\qquad{}+\{\omega_{2}\sin(\omega_{1}T)\cos(\omega_{2}T)
-\omega_{1}\cos(\omega_{1}T)\sin(\omega_{2}T)\}
\left\{\frac{\Pi_{z\,1}^{2}}{2}+\frac{\Pi_{z\,2}^{\ast\,2}}{2}\right\} \\
\qquad{}+\{\omega_{1}^{2}\omega_{2}\sin(\omega_{1}T)
+\omega_{1}\omega_{2}^{2}\sin(\omega_{2}T)\}\{x_{2}^{\ast}\Pi_{z\,1}+\Pi_{z\,2}^{\ast}x_{1}\} \\
 \qquad{} +\{-\omega_{1}^{2}\omega_{2}\sin(\omega_{1}T)\cos(\omega_{2}T)
 +\omega_{1}\omega_{2}^{2}\cos(\omega_{1}T)\sin(\omega_{2}T)\}
 \{x_{2}^{\ast}\Pi_{z\,2}^{\ast}+x_{1}\Pi_{z\,1}\} \\
\qquad{} +\{-\omega_{1}^{4}\omega_{2}\sin(\omega_{1}T)-\omega_{1}\omega_{2}^{4}
\sin(\omega_{2}T)\}\{x_{2}^{\ast}x_{1}\}\\
\qquad{}+\{-\omega_{2}\sin(\omega_{1}T)
-\omega_{1}\sin(\omega_{2}T)\}\{\Pi_{z\,2}^{\ast}\Pi_{z\,1}\}\Bigg\}\Biggr\},
\end{gather*}
or, for short:
\begin{gather}
 \langle x^{*}_{2}, \Pi_{z\,2}^{\prime*},t_{2} | x_{1}, \Pi_{z\,1},t_{1}\rangle = Q\left. \exp\left\{ \frac{i}{D}\right\{F(x_{2}^{*}\Pi_{z\,1}+x_{1}\Pi_{z\,2}^{*})
 \right. +G \Pi_{z\,1}\Pi_{z\,2}^{*}\nonumber\\
 \qquad{} +J \left(x_{2}^{*}\Pi_{z\,2}^{*}+x_{1}\Pi_{z\,1}\right)+K\left(\frac{\Pi_{z\,2}^{*2}+\Pi_{z\,1}^{2}}{2}\right) \left.\left. + M x_{1} x_{2}^{*}+ N \left(\frac{x_{2}^{*2}+x_{1}^{2}}{2}\right)\right\}\right\} , \label{PropPU}
\end{gather}
with $D$, $F$, $G$, $J$, $K$, $M$, $N$, $Q$, being the following functions of $T$:
\begin{gather*}
 D = (\omega_1^2-\omega_2^2)\sin(\omega_1T)\sin(\omega_2T), \\
 F =  \omega_1^2\omega_2\sin(\omega_1T)+\omega_1\omega_2^2\sin(\omega_2T), \\
 G = -\omega_2\sin(\omega_1T)-\omega_1\sin(\omega_2T), \\
 J = -\omega_1^2\omega_2\sin(\omega_1T)\cos(\omega_2T)+ \omega_1\omega_2^2 \sin(\omega_2T)\cos(\omega_1T), \\
 K = \omega_2\sin(\omega_1T)\cos(\omega_2T)-\!\! \omega_1 \sin(\omega_2T)\cos(\omega_1T) , \\
 M = -\omega_1^4 \omega_2\sin(\omega_1T)-\omega_1\omega_2^4\sin(\omega_2T), \\
 N = \omega_1^4 \omega_2\sin(\omega_1T)\cos(\omega_2T)- \omega_1\omega_2^4
 \sin(\omega_2T)\cos(\omega_1T) , \\
 Q = \sqrt{\frac{1}{\sin(\omega_{1}T)\sin(\omega_{2}T)}}.
\end{gather*}
Just as in our f\/irst example the PU propagator (\ref{PropPU}) can be related to the one corresponding to the Hamiltonian (\ref{HHO}) by using the change of basis
\begin{gather*}
  \langle P_{1},P_{2}|x, \Pi_z\rangle = \exp\left[(ax-b\Pi_z)P_{1}+(-icx+ib\Pi_z)P_{2}\right] .
\end{gather*}
The basis $|x,\Pi_z\rangle$ is complete with the measure
\begin{gather}\label{meas}
  d\mu_{_{PU}}=\frac{dx_R dx_I d\Pi_{zR} d\Pi_{zI}}{(2\pi)^2}\delta(b\Pi_{zR}-a x_R)\delta(b\Pi_{zI}-c x_I),
\end{gather}
where $x_R$, $x_I$, $\Pi_{zR}$, $\Pi_{zI}$ are real.

\subsection*{Acknowledgements}

This work was partially supported by the following grants:
CONACyT-SEP 51132F, CONACyT-SEP 47211-F, CONACyT-SEP 55310,
DGAPA-UNAM IN109107 and a CONACyT sabbatical grant to HAMT. AD
wishes also to acknowledge support from CONACyT.

\pdfbookmark[1]{References}{ref}
\LastPageEnding


\begin{thebibliography}{99}

\footnotesize\itemsep=0pt



\bibitem{Thirring:1950}
Thirring W.,
Regularization as a consequence of higher order equations,
{\it Phys. Rev.} {\bf 77} (1950), 570.

\bibitem{Pais:1950za}
Pais A., Uhlenbeck G.E.,
On f\/ield theories with nonlocalized action,
{\it Phys.\ Rev.} {\bf 79} (1950), 145--165.

\bibitem{Heisenberg1957}
Heisenberg W.,
Lee model and quantisation of non linear f\/ield theories,
{\it Nuclear Phys.} {\bf 4} (1957), 532--563.


\bibitem{Stelle:1976gc}
Stelle K.S.,
Renormalization of higher-derivative quantum gravity,
{\it Phys.\ Rev. D} {\bf 16} (1977), 953--969.

\bibitem{Tomboulis:1983sw}
Tomboulis E.T.,
Unitarity in higher derivative quantum gravity,
{\it Phys.\ Rev.\ Lett.} {\bf 52} (1984), 1173--1176.

\bibitem{Hawking:2001yt}
Hawking  S.W., Hertog T.,
Living with ghosts,
{\it Phys. Rev.~D} {\bf 65} (2002), 103515, 8~pages,
\href{http://arxiv.org/abs/hep-th/0107088}{hep-th/0107088}.

\bibitem{Moeller:2002vx}
Moeller N., Zwiebach B.,
Dynamics with inf\/initely many time derivatives and rolling tachyons,
{\it J. High Energy Phys.} {\bf 2002} (2002), no.~10, 034, 39~pages,
\href{http://arxiv.org/abs/hep-th/0207107}{hep-th/0207107}.

\bibitem{Rivelles:2003jd}
Rivelles V.O.,
Triviality of higher derivative theories,
{\it Phys.\ Lett.\ B} {\bf 577} (2003), 137--142,
\href{http://arxiv.org/abs/hep-th/0304073}{hep-th/0304073}.

\bibitem{Antoniadis:2006pc}
Antoniadis I., Dudas E., Ghilencea D.M.,
Living with ghosts and their radiative corrections,
{\it Nuclear Phys.~B} {\bf 767}(2007), 29--53, \href{http://arxiv.org/abs/hep-th/0608094}{hep-th/0608094}.

\bibitem{Codello:2006in}
Codello A., Percacci R.,
Fixed points of higher-derivative gravity,
{\it Phys.\ Rev.\ Lett.} {\bf 97} (2006), 221301, 4~pages,
\href{http://arxiv.org/abs/hep-th/0607128}{hep-th/0607128}.

\bibitem{Berkovits:2006vc}
Berkovits N.,
 New higher-derivative $R^4$ theorems,
{\it Phys.\ Rev.\ Lett.} {\bf 98} (2007), 211601, 4~pages,
\mbox{\href{http://arxiv.org/abs/hep-th/0609006}{hep-th/0609006}}.

\bibitem{Jaen:1986iz}
Jaen X., Llosa J., Molina A.,
A reduction of order two for inf\/inite order Lagrangians,
{\it Phys.\ Rev. D} {\bf 34} (1986), 2302--2311.

\bibitem{Eliezer:1989cr}
 Eliezer D.A., Woodard R.P.,
The problem of nonlocality in string theory,
{\it Nuclear Phys.\ B} {\bf 325} (1989), 389--469.

\bibitem{Cheng:2001du}
Cheng T.C., Ho P.M., Yeh M.C.,
Perturbative approach to higher derivative and nonlocal theories,
{\it Nuclear Phys.\ B} {\bf 625} (2002), 151--165,
\href{http://arxiv.org/abs/hep-th/0111160}{hep-th/0111160}.

\bibitem{Simon:1990ic}
Simon J.Z.,
Higher derivative Lagrangians, non-locality, problems and solutions,
{\it Phys.\ Rev.\ D} {\bf 41} (1990), 3720--3733.

\bibitem{Bender:2007wu}
Bender  C.M., Mannheim P.D.,
No-ghost theorem for the fourth-order derivative Pais--Uhlenbeck oscillator model,
{\it Phys.\ Rev.\ Lett.} {\bf 100} (2008), 110402, 4~pages,
\href{http://arxiv.org/abs/0706.0207}{arXiv:0706.0207}.


 \bibitem{Bender:2008vh1}
Bender C.M., Mannheim P.D.,
Exactly solvable ${\mathcal PT}$-symmetric Hamiltonian having no Hermitian counterpart,
{\it  Phys.\ Rev.\ D} {\bf 78} (2008), 025022, 20~pages,
\href{http://arxiv.org/abs/0804.4190}{arXiv:0804.4190}.


\bibitem{Bender:2008vh}
 Bender C.M., Mannheim P.D.,
Giving up the ghost,
{\it J.~Phys.~A: Math. Theor.} {\bf 41} (2008), 304018, 7~pages,
\href{http://arxiv.org/abs/0807.2607}{arXiv:0807.2607}.



\bibitem{Smilga:2004cy}
 Smilga A.V.,
Benign vs. malicious ghosts in higher-derivative theories,
{\it Nuclear Phys.\ B} {\bf 706} (2005), 598--614,
\href{http://arxiv.org/abs/hep-th/0407231}{hep-th/0407231}.

\bibitem{Smilga:2005gb}
Smilga A.V.,
Ghost-free higher-derivative theory,
{\it Phys.\ Lett.\ B} {\bf 632} (2006), 433--438,
\href{http://arxiv.org/abs/hep-th/0503213}{hep-th/0503213}.

\bibitem{Smilga1}
 Smilga A.V.,
Comments on the dynamics of the Pais--Uhlenbeck oscillator,
{\it SIGMA} {\bf 5} (2009), 017, 13~pages,
\href{http://arxiv.org/abs/0808.0139}{arXiv:0808.0139}.

\bibitem{Smilga:2008hn}
Smilga A.V.,
Exceptional points in quantum and classical dynamics,
{\it J.~Phys.~A: Math. Theor.} {\bf 42} (2009), 095301, 9~pages,
\href{http://arxiv.org/abs/0808.0575}{arXiv:0808.0575}.

\bibitem{Bender:2007nj}
Bender C.M.,
Making sense of non-Hermitian Hamiltonians,
{\it Rep. Progr. Phys.} {\bf 70} (2007), 947--1018,
\mbox{\href{http://arxiv.org/abs/hep-th/0703096}{hep-th/0703096}}.

\bibitem{Ashtekar:1991hf}
Ashtekar A.,
Lectures on nonperturbative canonical gravity,
{\it Advanced Series in Astrophysics and Cosmology}, Vol.~6,
World Scientif\/ic, Singapore, 1991.

\bibitem{Ashtekar:1992ne}
Ashtekar A.,
Mathematical problems of nonperturbative quantum general relativity,
\href{http://arxiv.org/abs/gr-qc/9302024}{gr-qc/9302024}.


\bibitem{Thiemann:1995ug}
Thiemann T.,
Reality conditions inducing transforms for quantum gauge f\/ield theory and quantum gravity,
{\it Classical Quantum Gravity} {\bf 13} (1996), 1383--1403,
\href{http://arxiv.org/abs/gr-qc/9511057}{gr-qc/9511057}.

\bibitem{Ashtekar:1995qw}
Ashtekar A.,
A generalized Wick transform for gravity,
{\it Phys.\ Rev.\ D} {\bf 53} (1996), 2865--2869,
\href{http://arxiv.org/abs/gr-qc/9511083}{gr-qc/9511083}.

\bibitem{Montesinos:1999qc}
Montesinos M., Morales-T\'ecotl H.A., Urrutia L.F., Vergara J.D.,
Complex canonical gravity and reality constraints,
{\it Gen. Relativity Gravitation} {\bf 31} (1999), 719--723.


\bibitem{Rovelli:2004tv}
Rovelli C.,
Quantum gravity,
Cambridge University Press, Cambridge, 2004.

\bibitem{Thiemann:2007zz}
Thiemann T.,
Modern canonical quantum general relativity,
Cambridge University Press, Cambridge, 2007.

\bibitem{OSTROGRADSKI} Ostrogradsky M.,
M\'emoires sur les \'equations dif\/f\'erentielles relatives aux probl\`emes des isop\'erim\`etres,
{\it Mem. Acad. St. Petersbourg} {\bf VI 4} (1850), 385--517.


\bibitem{Swanson}
Swanson M.S.,
Transition elements for a non-Hermitian quadratic Hamiltonian,
{\it J. \ Math. \ Phys.} {\bf 45} (2004), 585--601.

\bibitem{Jones}
Jones H.F., On pseudo-Hermitian Hamiltonians and their Hermitian counterparts,
{\it J.~Phys.~A: Math. Gen.} {\bf 38} (2005), 1741--1746,
\href{http://arxiv.org/abs/quant-ph/0411171}{quant-ph/0411171}.

\bibitem{Ivanov:2007me}
Ivanov E.A., Smilga A.V.,
Cryptoreality of nonanticommutative Hamiltonians,
{\it J. High Energy Phys.} {\bf 2007} (2007), no.~07, 036, 16~pages,
\href{http://arxiv.org/abs/hep-th/0703038}{hep-th/0703038}.


\bibitem{Anderson:1993ia}
Anderson A.,
Canonical transformations in quantum mechanics,
{\it Ann. Physics} {\bf 232} (1994), 292--331,
\mbox{\href{http://arxiv.org/abs/hep-th/9305054}{hep-th/9305054}}.


\bibitem{Anderson:1993im}
Anderson A.,
Quantum canonical transformations: physical equivalence of quantum theories,
{\it Phys.\ Lett.\ B} {\bf 305} (1993), 67--70,
\href{http://arxiv.org/abs/hep-th/9302062}{hep-th/9302062}.

\bibitem{Schwinger2001}
Schwinger J.,
Quantum mechanics, symbolism of atomic measurements,
Springer-Verlag, Berlin, 2001.

\bibitem{Henneaux}
Henneaux M., Teitelboim C.,
Quantization of gauge systems,
Princeton University Press, Princeton, 1992.

\bibitem{dirac}
 Dirac P.A.M.,
 Lectures on quantum mechanics,
 Belfast Graduate School of Science, New York, 1964.

\bibitem{Senjanovic}
 Senjanovic P.,
Path integral quantization of f\/ield theories with second class constraints,
{\it Ann. Physics} {\bf 100} (1976), 227--261,
Erratum, {\it Ann. Physics} {\bf 209} (1991), 248.


\bibitem{Mannheim1}
Mannheim  P.D., Davidson A.,
Dirac quantization of the Pais--Uhlenbeck fourth order oscillator,
{\it Phys.\ Rev.~A} {\bf 71} (2005), 042110, 9~pages,
\href{http://arxiv.org/abs/hep-th/0408104}{hep-th/0408104}.


\bibitem{Alexandrov:1998cu}
Alexandrov S.Y., Vassilevich D.V.,
Path integral for the Hilbert--Palatini and Ashtekar gravity,
{\it Phys.\ Rev.~D} {\bf 58} (1998), 124029, 13~pages,
\href{http://arxiv.org/abs/gr-qc/9806001}{gr-qc/9806001}.

\bibitem{Mannheim}
Mannheim P.D.,
Solution to the ghost problem in fourth order derivative theories,
{\it Found.\ Phys.} {\bf 37} (2007), 532--571,
\href{http://arxiv.org/abs/hep-th/0608154}{hep-th/0608154}.

\end{thebibliography}
\end{document}